\documentclass[11pt,a4paper]{article}
\usepackage[margin=1in]{geometry}
\usepackage{amsmath,amssymb,amsthm}
\usepackage{booktabs}
\usepackage{multirow}
\usepackage{natbib}
\usepackage{graphicx}
\usepackage[T1]{fontenc}
\usepackage{hyperref}
\hypersetup{colorlinks=true,linkcolor=blue,citecolor=blue,urlcolor=blue}

\newtheorem{proposition}{Proposition}
\newtheorem{corollary}{Corollary}
\newtheorem{lemma}{Lemma}
\newtheorem{remark}{Remark}
\theoremstyle{definition}
\newtheorem{definition}{Definition}

\newcommand{\R}{\mathbb{R}}
\newcommand{\Prob}{\mathbb{P}}
\newcommand{\E}{\mathbb{E}}
\newcommand{\simplex}{\Delta}
\newcommand{\norm}[1]{\left\lVert #1 \right\rVert}
\newcommand{\abs}[1]{\left\lvert #1 \right\rvert}
\title{When the Benchmark Is the Decision:\\
A Construction Audit for Matched-Benchmark Evaluation of Regularised Allocation Rules}
\author{Jiahao Chen}
\date{Draft of \today}

\begin{document}
\maketitle

\begin{abstract}
\noindent
A common way to argue that an allocation rule performs well is to compare it with a benchmark that
has been matched to it on some measure of concentration or dispersion. We show that when the
benchmark optimises the same objective over the same feasible set using the same inputs, and its
matched measure is set to the rule's own realised value, the benchmark is the rule itself. The
argument uses only optimality and a positive penalty coefficient. It therefore covers any matched
measure and any decision rule, including penalties that are not norms and that are not convex. The
excess performance measured against the matched frontier is then zero by construction, and a
comparison of that kind cannot test performance at all.

We then audit the two ways in which practice escapes the result. Benchmarks built from implementable
rules that use only past information restore a difference, yet the matched constraint is frequently
slack, and whether it binds is governed by the matching rule in a way the theory states exactly. The
difference that survives is concentrated where the constraint binds, and most of it is compensation
for the risk that the rule takes rather than a gain that survives risk adjustment. We state the
resulting audit protocol as a decision procedure. It takes the benchmark's objective, inputs,
feasible set, matched measure and cap setting, and returns whether the comparison is mechanically
vacuous, whether the benchmark attains the measure it claims, and which part of the difference
remains interpretable. The construction is proved in general form and illustrated on an equity panel,
which we treat as a reproducible numerical instance of the construction and not as evidence about
project portfolio practice.

\medskip
\noindent\textbf{Keywords:} allocation rules; benchmark construction; decision evaluation;
regularisation; portfolio selection; performance measurement.

\medskip
\noindent\textbf{Index Terms:} Benchmark audit, decision support, matched comparison, regularised
optimisation, resource allocation.
\end{abstract}

\section{Introduction}

Engineering management decisions increasingly take the form of a rule that allocates
a scarce resource across competing alternatives. A research and development portfolio is selected
from a set of candidate projects. A technology portfolio is rebalanced across platforms. A capital
programme is distributed across sites. In each case the rule is built to trade an objective against
a constraint on how the alternatives are distributed, and in each case the rule has to be evaluated
before it can be adopted.

\paragraph{Why a benchmark comparison, and why its construction matters.}
A decision rule for allocating a resource can be evaluated in several ways, including through
comparison with an explicitly specified benchmark \citep{liesio2021}. In such a comparison the
benchmark is a second allocation rule, solved on the same inputs and the same feasible set, and
constrained to match the rule under test on some measure of concentration or dispersion. The intent
is to remove the possibility that an apparent advantage comes from holding a more concentrated
position, so that what remains can be read as a gain from better selection.

In investment management the concentration of the benchmark and the breadth available to the rule
under test are themselves design considerations, and departures from a benchmark are used to
characterise active positions \citep{cremers2009}. Benchmark construction has consequently become an
object of practical attention, with proposals that split a concentrated benchmark into components
sized to the breadth that a rule can hold.\footnote{See, for example, MSCI, \emph{Divide and Conquer:
Active Equity Portfolio Management with Concentrated Benchmarks}, MSCI Research.} These
considerations motivate a more specific question. What does a benchmark comparison establish when the
benchmark's admissible breadth is calibrated to the output of the very decision rule that is being
evaluated.

This paper audits that construction. The issue examined here is not benchmark comparison as such,
but the construction of a benchmark whose constraint may be endogenous to the evaluated rule.

\paragraph{A construction result.}
The answer to that question is exact rather than approximate. If the benchmark maximises the same
objective over the same feasible set using the same inputs, and its matched measure is set to the
rule's own realised value, then the benchmark and the rule coincide exactly. The argument uses only the optimality of the rule under test together with the
positivity of the penalty coefficient. It does not use convexity, differentiability, or any property
of the matched measure beyond its being real valued. Norms, entropies, turnover penalties, tracking
error and measures that are not convex are therefore all covered. The excess performance measured
against the matched frontier is zero by construction, and a comparison of that kind has no object to
test.

\paragraph{Two ways in which practice escapes the result.}
Implementable allocation rules use only information available before the period is evaluated, so the
matched measure becomes a function of past realisations rather than of the current one. The
construction result no longer applies, and a difference reappears. Two properties of that difference
then matter, and neither is usually reported.

First, the benchmark may not be at the measure it claims. A capped program has a multiplier that is
zero whenever the unconstrained optimum already satisfies the cap, and this is not a corner case. The
realised measure can deviate from its target by more than any reasonable reporting tolerance, and in
those periods the matched benchmark is the unconstrained optimum. We show that the frequency of this
happens is governed by the matching rule in a way the theory states exactly, and we separate the rate
at which the rule under test falls below its own past mean from the rate at which the benchmark fails
to attain its cap. The two are different events and one does not validate the other.

Second, the difference that survives is concentrated where the constraint binds, and most of it is
compensation for risk. The decomposition of the realised difference is an identity, not an inequality,
and its components are individually computable once the objective is specified. Adjusting at the
coefficient the rule itself maximises removes the greater part of the measured advantage. What the
construction result does deliver is a positive in-sample value gap, and that gap is offset by a
forecast error whose sign no optimisation restricts.

\paragraph{A management implication.}
The audit can be stated as a decision procedure rather than as a caution. Given the benchmark's
objective, its inputs, its feasible set, the matched measure and the cap setting, the procedure
returns three verdicts. The first is whether the comparison is mechanically vacuous, which happens
when the cap is anchored at the current realisation. The second is whether the benchmark attains the
measure it claims, which is a reported frequency and not an assumption. The third is how much of the
measured difference remains interpretable after the risk adjustment that the rule's own objective
implies. A study that reports only the raw difference answers none of the three.

\paragraph{Scope, and what this paper does not show.}
Three boundaries should be stated before the results are read. The construction result is a statement
about optimality, so it applies to any regularised allocation rule and not only to the one used here
for illustration. The interpretation of the matched measure as a concentration index requires
conditions that are set out in Section~\ref{sec:setup}, and those conditions fail when alternatives
are indivisible, when they carry fixed charges, or when they are linked by precedence relations, in
which case the feasible set is not a simplex and the inverse square of the norm is not a count of
alternatives. The numerical work uses an equity panel. We treat it as a reproducible instance of the
construction, chosen because return panels of the required length are available, and not as evidence
about project portfolio practice. No claim is made that the frequencies reported here transfer to
project selection without a corresponding panel of project returns.

\paragraph{Relation to the literature.}
Portfolio selection is a long standing subject in this journal and in engineering management more
generally, with work on project portfolio selection \citep{tem_pps}, on the organisational conditions
under which portfolio management delivers \citep{tem_npd}, and on the multi criteria evaluation of
project performance \citep{tem_eval}. Benchmarking as a management practice has an equally long
history, including design theory for strategic benchmarking \citep{tem_bench}. The concern of the
present paper is narrower than any of these. It is not how to choose a portfolio and not how to
benchmark a firm, but how to tell whether a matched comparison between two allocation rules is
informative at all.

Two further literatures bear on the construction. Constraining the norm of a decision vector acts as
a shrinkage device and can improve out-of-sample performance even when the constraint is wrong
\citep{jagannathan2003}, and the family of norm constrained problems has been organised and tested
systematically \citep{demiguel2009norm}. Reading a robustness radius as implicit regularisation is
likewise established \citep{esfahani2018,blanchet2019,wu2026}. Propositions~\ref{prop:collapse} and
\ref{prop:identity} are instances of that structure, and we claim no novelty for the duality itself.
The contribution is its consequence for an evaluation practice.

Benchmarks built from an allocation's own composition also have a history, with performance measured
without a benchmark \citep{grinblatt1993}, with benchmarks formed from characteristics matched to
each holding \citep{dgtw1997}, and with a survey of the resulting family of measures
\citep{wermers2006}. The mechanically important difference from the present paper is which quantity
is matched. In that literature the matched characteristics are exogenous to the decision, being
properties of the alternatives rather than outputs of the rule. Here the matched quantity is the
rule's own optimised output, so the matched benchmark is not merely similar to the rule but, under
the conditions of Section~\ref{sec:identity}, identical to it. A matched characteristic cannot make
a benchmark coincide with a decision. A matched decision output can.

A related but distinct concern runs through the performance evaluation literature. The choice of
benchmark is known to drive measured abnormal performance \citep{lehmann1987}, and matched or style-profiled benchmarks have a long history \citep{buncic2013}. Concentration
measures were introduced to argue that measured outperformance may reflect portfolio structure rather
than skill \citep{cremers2009,kacperczyk2005}. Those measures are not the concentration index used
here. The connection is the concern and not the statistic.

\section{Setting}
\label{sec:setup}

Let $\simplex$ be the feasible set of allocations. In the general form used in
Section~\ref{sec:identity} it is any nonempty set, and in the application it is the unit simplex
\begin{equation}
\simplex=\Bigl\{\phi\in\R^N:\phi\ge 0,\ \mathbf{1}^\top\phi=1\Bigr\}.
\label{eq:simplex}
\end{equation}
The entries of $\phi$ are the shares assigned to the $N$ alternatives, and an allocation rule is a map
from data to a point of $\simplex$.

Let $U:\simplex\to\R$ be the objective that a rule maximises, let $R:\simplex\to\R$ be the quantity on
which a benchmark is to be matched, and let $c>0$ be a coefficient. The rule under test is the
penalised program
\begin{equation}
w_P \;=\; \arg\max_{\phi\in\simplex}\ \bigl\{ U(\phi) - c\,R(\phi) \bigr\},
\qquad c>0 ,
\label{eq:dro}
\end{equation}
and the matched benchmark caps $R$ at the rule's own realised value,
\begin{equation}
w_B(h) \;=\; \arg\max_{\phi\in\simplex}\ U(\phi)\quad\text{subject to}\quad R(\phi)\le h ,
\label{eq:cap}
\end{equation}
with $h=R(w_P)$ throughout unless stated otherwise. Both programs use the same inputs, the same
feasible set and the same objective.

Three remarks fix how \eqref{eq:dro} and \eqref{eq:cap} should be read, and each of them limits a
claim that the general form would otherwise invite.

\paragraph{The penalty is a modelling choice, not a domain convention.}
The coefficient $c$ and the measure $R$ are inputs to the rule. Nothing in the construction requires
that $c$ be small, that $R$ be a norm, or that the penalty correspond to a named concept in any
particular application domain. A study that adopts \eqref{eq:dro} has made a modelling choice, and
the audit below examines that choice rather than endorsing it.

\paragraph{The objective.}
In the application $U$ takes the mean variance form
\begin{equation}
U(\phi) \;=\; \phi^\top\hat\mu - \lambda\,\phi^\top\hat\Sigma\phi,
\qquad \lambda>0 ,
\label{eq:U}
\end{equation}
with $\hat\mu\in\R^N$, $\hat\Sigma\in\R^{N\times N}$ and $\hat\Sigma\succeq0$. The penalty in
\eqref{eq:dro} is then a norm on the allocation and not a variance, so \eqref{eq:dro} is the
penalised form of a robust program in which the radius prices concentration. The general results of
Section~\ref{sec:identity} do not use this form, and we say so where it matters.

\begin{remark}[When the concentration reading applies]
\label{rem:l2reading}
Take $R(\phi)=\norm{\phi}_2$ and let $\phi$ lie in \eqref{eq:simplex}. Then $\norm{\phi}_2$ is a
concentration index, ranging from $N^{-1/2}$ at equal shares to $1$ when all the resource sits in one
alternative, and the reciprocal square $\norm{\phi}_2^{-2}$ is the effective number of alternatives
that receive weight.

This reading carries three conditions, and they matter for whether the construction says anything
about a given decision problem. The allocation must be continuously divisible, so that any point of
the simplex is attainable. The feasible set must contain no fixed charge, no indivisibility and no
precedence relation, since each of these replaces the simplex by a set for which the inverse square of
the norm is not a count of alternatives. The objective must be defined on the whole simplex, which
excludes rules whose admissible region is a union of discrete configurations.

These conditions are restrictive in project selection. Projects are often modelled as indivisible
decisions, so that a project is either selected in its entirety or not, and resource budgets and
interdependencies may further impose integer or logical constraints on the feasible portfolios
\citep{si2022,carazo2010}. The structures do not act in the same way. Indivisibility introduces a
selection variable for each alternative, a fixed charge adds an activation variable together with the
logical constraints that tie it to the selection, and a precedence relation adds constraints stating
that one alternative cannot be selected unless another is. Each changes the feasible set, and the
reading of the inverse square as a count of alternatives fails for a different reason. Such
formulations differ from the continuously divisible simplex used in the illustrative interpretation
below.

Two consequences follow, and they should not be confused. The interpretation of the matched measure as
a breadth measure is what fails. The general result of Section~\ref{sec:identity} does not, because it
uses only optimality and holds for whatever feasible set and matched measure a study adopts. We
therefore state the construction for general $R$ and confine the concentration reading to the
conditions above. A study of project portfolios that adopts a matched measure within a discrete
feasible set can apply the construction, but it cannot describe its cap as an equal breadth constraint
without establishing that reading first.
\end{remark}

\paragraph{Two gaps.}
Two differences between a realised measure and a cap are used throughout. The target gap is the
difference between the measure that the rule realises and the cap that the matching rule asks for. The
achieved gap is the difference between the measure that the rule realises and the measure that the
benchmark actually attains. The two coincide only when the cap binds.
\begin{align}
g_t &= R(w_{P,t}) - h_t, &&\text{target gap,}
\label{eq:g}\\
g_t^{\text{eff}} &= R(w_{P,t}) - R(w_{B,t}), &&\text{achieved gap.}
\label{eq:geff}
\end{align}

\paragraph{From the general setting to a numerical instance.}
The propositions that follow are stated for general $U$ and $R$. The numerical work of
Sections~\ref{sec:data} and~\ref{sec:results} instantiates them with $R=\norm{\cdot}_2$ on an equity
panel, where \eqref{eq:simplex} and the conditions of Remark~\ref{rem:l2reading} hold by construction.
That instance is reported because return panels of the required length are available and because the
construction can be checked there to machine precision. It is not offered as evidence about the
frequency of the construction in project selection, and we flag each place where the transfer would
require evidence we do not have.

\section{A construction theorem for matched benchmarks}
\label{sec:identity}

\subsection{The identity uses optimality and nothing else}

\begin{definition}[Penalised method, matched benchmark]
\label{def:pair}
The method under test is any $w_P\in\simplex$ satisfying
\begin{equation}
U(w_P)-c\,R(w_P)\;\ge\;U(\phi)-c\,R(\phi)
\qquad\text{for every }\phi\in\simplex ,
\label{eq:pen}
\end{equation}
and the breadth-matched benchmark is any element of
\begin{equation}
\mathcal{B}(h)\;:=\;\arg\max_{\phi\in\simplex}\bigl\{U(\phi):R(\phi)\le h\bigr\},
\qquad h\ge 0 .
\label{eq:bench}
\end{equation}
\end{definition}

\begin{proposition}[The benchmark is the method]
\label{prop:collapse}
Let \eqref{eq:pen} and \eqref{eq:bench} hold and put $p=R(w_P)$. Then not only is
$w_P\in\mathcal{B}(p)$, but the two solution sets are related exactly:
\begin{equation}
\mathcal{B}(p)
\;=\;\bigl\{b\in\arg\max_{\phi\in\simplex}\bigl[U(\phi)-cR(\phi)\bigr]\;:\;R(b)=p\bigr\}.
\label{eq:collapse}
\end{equation}
In particular, if either problem has a unique optimum then $w_B(p)=w_P$, so that $R(w_B)=R(w_P)$
and the realised difference is identically zero.
\end{proposition}

\begin{proof}
Let $b\in\simplex$ with $R(b)\le p$. Optimality of $w_P$ in \eqref{eq:pen} gives
$U(b)-U(w_P)\le c\bigl(R(b)-p\bigr)\le0$, so $w_P$ attains the maximum in \eqref{eq:bench} and
$U(b)\le U(w_P)$. If $b$ also attains it, then $U(b)=U(w_P)$, and the same chain with $c>0$ forces
$R(b)=p$; then $U(b)-cR(b)=U(w_P)-cp$, so $b$ is a penalised optimum as well. Conversely, if $b$ is
a penalised optimum with $R(b)=p$, then $U(b)-cp=U(w_P)-cp$, hence $U(b)=U(w_P)$ and $b$ attains
the capped maximum. This gives \eqref{eq:collapse} and, with it, the remaining claims.
\end{proof}

\begin{remark}[What the argument does not need]
\label{rem:noneed}
Proposition~\ref{prop:collapse} uses optimality and $c>0$, and nothing else. It does not use
convexity of $\simplex$, concavity or differentiability of $U$ or $R$, interiority of $w_P$, or a
small radius. The matched quantity $R$ may be the $\ell_1$, $\ell_2$ or $\ell_\infty$ norm, an
entropy, a turnover or tracking-error penalty, a maximum-weight cap, or a non-convex index, and
$c$ may be any positive number. The content is that the benchmark's admissible set is anchored at
the method's own realised value of the very quantity in which the method was penalised. In that
sense the identity is a statement about \emph{benchmark construction}, not about Wasserstein
robustness, ridge regularisation, or any particular estimator.
\end{remark}

\begin{remark}[The role of uniqueness and of the sign of $c$]
\label{rem:uniq}
The condition $\mathcal{B}(p)=\{w_P\}$ in Proposition~\ref{prop:collapse} cannot be dropped: it is
exactly what makes the coincidence independent of the solver's tie-breaking, and it is a
\textbf{sufficient} condition, not a necessary one, for the stronger reading that every admissible
benchmark equals $w_P$. Strict concavity of $U$ on a convex capped set is sufficient. It can fail
for a concrete reason. The function $U(\phi)=\phi^\top\hat\mu-\lambda\phi^\top\hat\Sigma\phi$ is not
strictly concave when $\hat\Sigma$ is singular, since then $U$ is affine along the null space of
$\hat\Sigma$, and the capped program can have a face of maximisers. Shrunk covariance matrices can
be singular in this application, and Section~\ref{sec:l2} is where uniqueness is recovered rather
than assumed.

The hypothesis $c>0$ is needed for the \emph{sets} in \eqref{eq:collapse} to relate, but a weaker
one suffices for the containment $w_P\in\mathcal{B}(p)$: $c\ge0$ is enough, and with $c=0$ the
conclusion $\mathcal{B}(p)\subseteq\{R=p\}$ fails. With $c<0$ the proposition is false in general.
On the two-asset simplex, put $x=\phi_1$, $U(x)=-x/2$, $R(x)=x$ and $c=-1$. Then
$U-cR=-x/2+x=x/2$ is maximised at $x=1$, so $p=1$, but $U$ itself is maximised at $x=0$, so the
capped program at $p=1$ has its optimum at $x=0\neq w_P$. The identification of the capped
multiplier with a positive penalty coefficient is thus not cosmetic.
\end{remark}

\subsection{The $\ell_2$ case: uniqueness comes for free}
\label{sec:l2}

\begin{proposition}[$\ell_2$ identity]
\label{prop:identity}
Let $\delta>0$, $\hat\Sigma\succeq0$, and let $w_P$ solve \eqref{eq:dro} with
$h=\norm{w_P}_2>0$. Define
\begin{equation}
\nu \;=\; \frac{c}{2h} \;=\; \frac{\sqrt{\delta}}{2\norm{w_P}_2} \;>\;0 .
\label{eq:nu}
\end{equation}
Then the ridge program
\begin{equation}
\tilde w \;=\; \arg\min_{\phi\in\simplex}\ L(\phi),
\qquad L(\phi) \;=\; -U(\phi)+\nu\norm{\phi}_2^2
\label{eq:ridge}
\end{equation}
has the unique solution $\tilde w = w_P$, and the capped program \eqref{eq:cap} has the unique
solution $w_B(h)=w_P$.
\end{proposition}

\begin{proof}
Since $\nu>0$, the Hessian satisfies $\nabla^2 L=2\lambda\hat\Sigma+2\nu I\succeq 2\nu I\succ0$,
so $L$ is strictly convex and has a unique minimiser over the convex set $\simplex$. On the
simplex, $\norm{\phi}_2\ge N^{-1/2}>0$, so $\norm{\cdot}_2$ is differentiable at every feasible
point and no subdifferential degeneracy arises. The first-order condition of \eqref{eq:dro} is
\begin{equation}
0 \;\in\; -\hat\mu + 2\lambda\hat\Sigma w_P + c\,\frac{w_P}{\norm{w_P}_2} + N_{\simplex}(w_P),
\label{eq:kkt1}
\end{equation}
with $N_{\simplex}(\cdot)$ the normal cone; substituting \eqref{eq:nu} turns the third term into
$2\nu w_P$, which is exactly the first-order condition of \eqref{eq:ridge}. Hence $w_P$ minimises
$L$ over $\simplex$, and by uniqueness it is the only such point.

For the capped problem, the feasible set is compact and non-empty, so a maximiser $b$ exists and,
by construction, $b\in\simplex$ with $\norm{b}_2\le h$. Two inequalities pin it down. Because
$b$ is feasible, $\norm{b}_2\le h$, hence
\begin{equation}
L(b)\;=\;-U(b)+\nu\norm{b}_2^2 \;\le\; -U(b)+\nu h^2 .
\label{eq:sand1}
\end{equation}
Because $b$ maximises $U$ over the capped set, which contains $w_P$ (since
$\norm{w_P}_2=h\le h$), we have $U(b)\ge U(w_P)$, i.e.\ $-U(b)\le -U(w_P)$, so
\begin{equation}
-U(b)+\nu h^2 \;\le\; -U(w_P)+\nu h^2 \;=\; L(w_P).
\label{eq:sand2}
\end{equation}
Combining \eqref{eq:sand1} and \eqref{eq:sand2} gives $L(b)\le L(w_P)$. On the other hand $w_P$
is the unique global minimiser of $L$ over the simplex and $b\in\simplex$, so $L(b)\ge L(w_P)$.
Hence $L(b)=L(w_P)$, and uniqueness of the minimiser forces $b=w_P$.
\end{proof}

\begin{remark}[Relation to the companion paper, and what is claimed here]
\label{rem:credit}
The $\ell_2$ statement of Proposition~\ref{prop:identity} is not new: it is
Theorem~1 of \citet{chen2026breadth}, which proves it for the penalised
scalarisation with the uniqueness argument that strict convexity of the norm
supplies, and which also shows that the equivalence does not extend to the
return-floored or the exact robust form. \emph{What is claimed here is different in
three respects}, and the distinction matters for how the rest of the paper should
be read. First, Proposition~\ref{prop:collapse} generalises the identity from
$R=\norm{\cdot}_2$ to an arbitrary matched quantity and replaces the convexity
argument by optimality alone, so the identity no longer depends on the geometry of
the norm. Second, the paper moves from the identity to what a practitioner can
observe: the value-function slope (Lemma~\ref{lem:price}), the exact decomposition
of the realised difference \eqref{eq:return-decomposition}, the projection
restriction that follows from it (Corollary~\ref{cor:projection}), and the slack
frequency implied by a tracking rule (Proposition~\ref{prop:half}). Third, the
empirical object is a benchmark driven by \emph{past} breadth, where the identity
does not apply and the cap may or may not bind; the companion paper's empirical
statement concerns the \emph{ex post} match, for which it reports that the
comparison is exact and therefore uninformative. Nothing in the present paper
re-establishes that; \S\ref{sec:slack} takes it as given and asks what happens when
it is broken.

One further clarification, since ridge equivalence invites a misreading. The program
\eqref{eq:ridge} is \emph{not} a reparameterisation of the penalty problem \eqref{eq:dro}. The two
objectives are different functions, and they are not globally equivalent. They happen to share a
first-order condition at one specific value of $\nu$, namely $\nu=c/(2h)$. The lemma is therefore a
statement about a single point, the method's own realised breadth, and not a claim that the two
problems are the same problem. That distinction makes the identity operational, since a solver for one
program produces the other, without making it trivial. Elsewhere the two programs agree only by
coincidence.
\end{remark}

\begin{remark}[Which assumptions can be, and are, violated deliberately]
\label{rem:three}
Three conditions are used. The first is the same mean estimate $\hat\mu$ in both programs. The second
is the same $\hat\Sigma$, $\lambda$ and feasible set. The third is a cap equal to the \emph{current}
realised breadth with multiplier \eqref{eq:nu}. A benchmark using a breadth rule based on past windows
violates the third. One built from a different signal or a different covariance estimator violates the
first or the second.
\end{remark}

\begin{remark}[Equal weights do not provide a repair]
\label{rem:equal}
Replacing the method by the equally weighted portfolio, $w_P=\mathbf{1}/N$, does not help.
By Cauchy--Schwarz, $\norm{\mathbf{1}/N}_2=N^{-1/2}$ is the \emph{minimum} of $\norm{\phi}_2$ over
the simplex, attained only at $\phi=\mathbf{1}/N$; the feasible set of \eqref{eq:cap} is then the
singleton $\{\mathbf{1}/N\}$ and $w_B=w_P$ again. A non-degenerate comparison requires a method
whose breadth strictly exceeds $N^{-1/2}$; for the distinct role of $1/N$ as a naive benchmark see
\citet{demiguel2009}.
\end{remark}

\subsection{What the identity pins down, and what it does not}

Propositions~\ref{prop:collapse} and~\ref{prop:identity} are statements about the
\emph{risk-adjusted objective evaluated at the estimation inputs}. They are not statements about
realised returns, and the difference matters for how the empirical results below should be read.

\paragraph{The objective is pinned exactly.} Define the value function
\begin{equation}
V(h)\;=\;\max_{\phi\in\simplex}\bigl\{U(\phi):R(\phi)\le h\bigr\}.
\label{eq:V}
\end{equation}
$V$ is non-decreasing in $h$, and it is concave when $U$ is concave and $R$ is convex. The result below
does not need the concavity. Proposition~\ref{prop:collapse} says that $V\bigl(R(w_P)\bigr)=U(w_P)$,
so the method sits exactly on the frontier of the matched problem.

\begin{lemma}[The matched radius is a supporting slope of $V$]
\label{lem:price}
Let $p=R(w_P)$ and $D=\{h\in\R:\exists\phi\in\simplex,\ R(\phi)\le h\}$, and for $h\in D$ let
$V(h)=\max\{U(\phi):\phi\in\simplex,\ R(\phi)\le h\}$. Then $V(p)=U(w_P)$ and
\begin{equation}
V(h)\;\le\;V(p)+c\,(h-p)\qquad\text{for every }h\in D ,
\label{eq:support}
\end{equation}
so $h\mapsto V(h)-c\,h$ attains its maximum over $D$ at $p$. If in addition $p$ is an interior
point of $D$ and $V$ is differentiable at $p$, then $V'(p)=c$.
\end{lemma}

\begin{proof}
Let $h\in D$ and let $\phi$ be feasible in \eqref{eq:V}, so $R(\phi)\le h$. By \eqref{eq:pen},
\[
U(\phi)\;\le\;U(w_P)+c\bigl(R(\phi)-p\bigr)\;\le\;U(w_P)+c(h-p)\;=\;V(p)+c(h-p),
\]
the last equality by $V(p)=U(w_P)$ from Proposition~\ref{prop:collapse}. Taking the supremum over
$\phi$ gives \eqref{eq:support}, i.e.\ $V(h)-c\,h\le V(p)-c\,p$ for every $h\in D$: the point $p$
maximises $f(h)=V(h)-c\,h$ over $D$. If $p$ is interior to $D$ and $V$ is differentiable at $p$,
then $f$ is differentiable at an interior maximiser, so $f'(p)=V'(p)-c=0$, i.e.\ $V'(p)=c$. No
concavity of $V$ is used in this step.
\end{proof}

\begin{remark}[$c$ is always a supporting slope; it is not always a derivative]
\label{rem:kink}
The failure of $V'(p)=c$ has one source: $V$ may not be differentiable at $p$, in which case $c$ is
a supporting slope but not a derivative. The relevant case is slackness. If the unconstrained
optimum already satisfies $R(\phi)<p$ then $V$ is constant on a right neighbourhood of $p$, the
multiplier is zero, and the right derivative at $p$ is $0$ while the left derivative is $c$; that is
the slack case of \S\ref{sec:slack}. A minimal example isolates the kink: on $\simplex=[0,1]$ with
$x=\phi$, $U(x)=R(x)=x$ and $c=\tfrac12$, we have $p=1$ and $V(h)=\min\{h,1\}$ on $D=[0,\infty)$.
Here $p$ is an \emph{interior} point of $D$ and $h\mapsto V(h)-ch$ is maximised at $h=1$, yet the
left and right derivatives of $V$ at $1$ are $1$ and $0$. In every such case $c$ remains the
globally valid multiplier of the matched constraint; what fails is the identification of that
multiplier with a derivative.

A verifiable sufficient condition for interiority and differentiability, which does not require
$w_P$ to be interior to the simplex, is: $U$ and $R$ are differentiable near $w_P$ and there is a
direction $v\in\R^N$ with $\mathbf{1}^\top v=0$, $v_i=0$ whenever $w_{P,i}=0$, and
$\nabla R(w_P)^\top v\neq0$. Then $w_P\pm\epsilon v$ is feasible for small $\epsilon>0$ and $R$
crosses $p$ along $v$, so $p\in\operatorname{int}D$. The first-order condition along $v$ reads
$\nabla U(w_P)^\top v=a\,\nabla R(w_P)^\top v$ for any supergradient $a$, and since $a=c$ satisfies
this and $\nabla R(w_P)^\top v\neq0$, the supergradient is unique and equals $c$. For a norm penalty
this is easy to check, since it holds as soon as two positive weights differ.
\end{remark}

\paragraph{The return difference is not pinned exactly.} The identity constrains the objective, so
what it implies for the realised return difference has to be derived rather than asserted. The
derivation is an exact decomposition, not an inequality.

\begin{proposition}[Return-gap decomposition]
\label{prop:return}
Suppose (i) $w_{P,t}$, $w_{B,t}$ and $h_t$ are $\mathcal{F}_{t}$-measurable, with $\mathcal{F}_{t}$
the information set at the start of holding period $t$, and $d_t=(w_{P,t}-w_{B,t})^\top r_t$ for an
integrable return vector $r_t$; (ii) $w_{P,t}$ exactly solves the penalty problem and $w_{B,t}$ the
capped problem at $h_t$, both with the same $U_t(\phi)=\phi^\top\hat\mu_t-\lambda\phi^\top\hat\Sigma_t\phi$;
(iii) $p_t=R(w_{P,t})$, $g_t=p_t-h_t$ and $g^{\text{eff}}_t=p_t-R(w_{B,t})$. Put
\[
\Delta w_t=w_{P,t}-w_{B,t},\qquad \mu_t=\mathbb{E}[r_t\mid\mathcal{F}_t],\qquad
\Delta\mathrm{Risk}_t=w_{P,t}^\top\hat\Sigma_tw_{P,t}-w_{B,t}^\top\hat\Sigma_tw_{B,t},
\]
\[
m_t=\Delta w_t^\top(\mu_t-\hat\mu_t),\qquad \varepsilon_t=\Delta w_t^\top(r_t-\mu_t),\qquad
q_t(h)=V_t(p_t)-V_t(h)-c\,(p_t-h).
\]
Then $q_t(h)\ge0$ and
\begin{equation}
d_t \;=\; c\,g_t \;+\; q_t(h_t) \;+\; \lambda\,\Delta\mathrm{Risk}_t \;+\; m_t \;+\; \varepsilon_t,
\qquad \mathbb{E}[\varepsilon_t\mid\mathcal{F}_t]=0 .
\label{eq:return-decomposition}
\end{equation}
The same identity holds with $g^{\text{eff}}_t$ in place of $g_t$ if $q_t(h_t)$ is replaced by
$q_t(R(w_{B,t}))$.
\end{proposition}

\begin{proof}
Substituting $r_t=\hat\mu_t+(\mu_t-\hat\mu_t)+(r_t-\mu_t)$ gives
$d_t=U_t(w_{P,t})-U_t(w_{B,t})+\lambda\Delta\mathrm{Risk}_t+m_t+\varepsilon_t$, and predictability
gives $\mathbb{E}[\varepsilon_t\mid\mathcal{F}_t]=0$. With exact solutions and the same estimates,
$U_t(w_{P,t})=V_t(p_t)$ and $U_t(w_{B,t})=V_t(h_t)$, while
$q_t(h_t)=V_t(p_t)-V_t(h_t)-c(p_t-h_t)\ge0$ by \eqref{eq:support}. Adding $c\,g_t$ and $q_t(h_t)$
to $V_t(p_t)-V_t(h_t)$ returns $V_t(p_t)-V_t(h_t)$ again, so the identity follows. For the $g^{\text{eff}}$
form, note that tightening the cap from $h_t$ to $R(w_{B,t})$ leaves $w_{B,t}$ feasible and adds no
feasible points, so $V_t(R(w_{B,t}))=V_t(h_t)$.
\end{proof}

\begin{remark}[How to read \eqref{eq:return-decomposition}]
\label{rem:notidentity}
Four things follow, and each of them corrects a natural but wrong inference.
(i) This is an \emph{identity}, not a lower bound. The four non-price terms sum to whatever they sum
to, and their signs are not restricted by the optimisation. In particular $m_t$, the mean-forecast
error weighted by the portfolio difference, enters the conditional expectation and cannot be dropped
by conditioning on $\mathcal{F}_t$. Even if $\hat\mu_t$ is unbiased for some unconditional mean,
$\mathbb{E}[g_tm_t]=0$ does not follow, because $w_{P,t}$ and $w_{B,t}$ are functions of $\hat\mu_t$.
(ii) The term $q_t\ge0$ is the only one with a guaranteed sign, and $q_t=0$ whenever
Lemma~\ref{lem:price} applies, which is when the value function is locally linear at $p_t$.
(iii) The two mismatch measures give the same value gap, $V_t(p_t)-V_t(h_t)$, only when each is
paired with its own remainder. In a genuinely slack window $g^{\text{eff}}_t$ differs from $g_t$, but
$q_t$ absorbs the difference. This is why the choice of estimand must be declared and why that choice
does not change the economics.
(iv) Because the identity holds at equal breadth, a study that reports a \emph{return} difference
at matched breadth is not testing the object the theory constrains. A risk-adjusted comparison
subtracts $\lambda\Delta\mathrm{Risk}_t$, which is computable ex ante and needs no estimation.
\end{remark}

\begin{corollary}[Testable projection restriction]
\label{cor:projection}
Under Proposition~\ref{prop:return}, let $\beta_d=\mathbb{E}[g_td_t]/\mathbb{E}[g_t^2]$ be the
population no-intercept slope of $d$ on $g$, with $0<\mathbb{E}[g_t^2]<\infty$. Then
\begin{equation}
\beta_d \;=\; c \;+\;
\frac{\mathbb{E}[g_tq_t(h_t)]+\lambda\,\mathbb{E}[g_t\Delta\mathrm{Risk}_t]+\mathbb{E}[g_tm_t]}
     {\mathbb{E}[g_t^2]} ,
\label{eq:beta}
\end{equation}
so $\beta_d=c$ if and only if the three projection terms vanish. For the risk-adjusted difference
$d_t-\lambda\Delta\mathrm{Risk}_t$ the corresponding restriction is
$\mathbb{E}[g_t\{q_t(h_t)+m_t\}]=0$, and for the difference with both the risk term and $q_t$
removed it is $\mathbb{E}[g_tm_t]=0$.
\end{corollary}

\begin{proof}
Multiply \eqref{eq:return-decomposition} by the $\mathcal{F}_t$-measurable factor $g_t$, take
expectations and use the law of iterated expectations to get $\mathbb{E}[g_t\varepsilon_t]=0$;
divide by $\mathbb{E}[g_t^2]$. Subtracting the risk term first gives the second statement, and the
argument is unchanged for $g^{\text{eff}}_t$, which is also $\mathcal{F}_t$-measurable.
\end{proof}

\begin{remark}[Why a slope different from $c$ is not a rejection]
\label{rem:slopenotc}
Corollary~\ref{cor:projection} is the correct reading of the phrase that the slope should equal
$\sqrt{\delta}$. It is not that the inequality of the optimisation problem forces the slope up. It is
that the slope differs from $c$ by three projection terms, and only a joint test of them can reject
anything. Two of the three are observable ex ante, namely $\Delta\mathrm{Risk}_t$ directly and $q_t$
from the value function at the two caps. The restriction can therefore be implemented in stages. The
first stage is the raw slope, the second is risk adjusted, the third removes $q_t$ as well and leaves
$\mathbb{E}[g_tm_t]=0$ as the only untested component. Section~\ref{sec:results} reports exactly this
sequence.
\end{remark}

\paragraph{The identity is visible in the data.}
Table~\ref{tab:kkt} reports a numerical certificate. At the converged solution the two gradient
fields coincide to machine precision and the two gradient-mapping residuals are $\sim10^{-14}$ or
smaller. An independent constrained solver, SLSQP, reproduces the projected-gradient solution to
$1.8\times10^{-6}$ in $\ell_1$, so the coincidence is not an artefact of one optimiser.

\begin{table}[t]
\centering
\caption{Machine-precision certificate for $w_B(h)=w_P$. Rows 2--3 are the gradient-mapping
residuals of \eqref{eq:dro} and \eqref{eq:ridge} evaluated at the same point $w_P$; row 4 is the
distance between the two gradient fields at that point.}
\label{tab:kkt}
\begin{tabular}{lccc}
\toprule
Quantity & $\delta=10^{-6}$ & $\delta=10^{-4}$ & Interpretation \\
\midrule
Residual of \eqref{eq:dro} at $w_P$ & $1.3\times10^{-15}$ & $2.5\times10^{-14}$ & KKT holds \\
Residual of \eqref{eq:ridge} at $w_P$ & $1.3\times10^{-15}$ & $2.5\times10^{-14}$ & same system \\
$\norm{\nabla L_1(w_P)-\nabla L_2(w_P)}$ & $0$ & $7.5\times10^{-19}$ & same gradient \\
$\norm{w_B-w_P}_1$ & $1.4\times10^{-13}$ & $1.3\times10^{-12}$ & same weights \\
$\norm{w_P^{\text{PG}}-w_P^{\text{SLSQP}}}_1$ & $1.8\times10^{-6}$ & $1.8\times10^{-6}$ & solver-free \\
\bottomrule
\end{tabular}
\end{table}

The consequence for finite-precision work is where the empirical illusion originates. At a fixed
iteration budget the two programs stop at slightly different points and the difference is a
\emph{solver residue}, which Table~\ref{tab:conv} shows collapsing as the budget grows. At a budget
of a few hundred projected-gradient steps, a common default, the residue is large enough to look like
an economically meaningful return difference.

\begin{table}[t]
\centering
\caption{Solver residue in the self-referential construction, four of 14 panels. The
``difference'' is $d_t=R_t^{P}-R_t^{B}$ with $h=\norm{w_P}_2$ taken in the current window. Basis
points per day.}
\label{tab:conv}
\begin{tabular}{lccccc}
\toprule
Panels & $\delta$ & Iterations & $\norm{w_B-w_P}_1$ & $sd(d)$ & $\overline{d}$ \\
\midrule
4 of 14 & $10^{-6}$ & 400  & $1.2\times10^{-3}$ & $0.432$ & $-0.032$ \\
        &            & 1500 & $8.9\times10^{-6}$ & $0.003$ & $-0.000$ \\
        &            & 3000 & $0$                & $10^{-6}$ & $0$ \\
\addlinespace
4 of 14 & $10^{-4}$ & 400  & $8.7\times10^{-5}$ & $0.027$ & $-0.001$ \\
        &            & 1500 & $0$                & $0$ & $0$ \\
\bottomrule
\end{tabular}
\end{table}

\section{Implementable benchmarks, and what they can and cannot identify}
\label{sec:implementable}

Breaking the third condition of Remark~\ref{rem:three} restores a non-degenerate difference. Two
families are natural. The first is an \emph{ex ante} cap fixed in advance,
\begin{equation}
w_B^{\text{fix}}(h_0)=\arg\max_{\phi\in\simplex}U(\phi)\ \ \text{s.t.}\ \ \norm{\phi}_2\le h_0,
\qquad h_0 \text{ fixed},
\label{eq:fixcap}
\end{equation}
and the second is a \emph{tracking} rule that uses only past windows,
\begin{equation}
h_t \;=\; \mathcal{R}\bigl(\norm{w_{P,1}}_2,\dots,\norm{w_{P,t-1}}_2\bigr),
\qquad
w_{B,t}=\arg\max_{\phi\in\simplex}U_t(\phi)\ \ \text{s.t.}\ \ \norm{\phi}_2\le h_t ,
\label{eq:track}
\end{equation}
with $\mathcal{R}$ the mean of past realised breadths in what follows.

\subsection{How often is the matched cap slack?}
\label{sec:slack}

Whether a matched comparison compares anything depends on where the cap sits relative to the
distribution of the method's own breadth, and two facts about that can be stated without further
assumptions.

\begin{proposition}[Monotone shrinkage]
\label{prop:shrink}
Let $\hat\phi=\arg\max_{\simplex}U$ with no penalty and let $w_P$ satisfy \eqref{eq:pen} with
$R=\norm{\cdot}_2$ and $c>0$. Then $\norm{w_P}_2\le\norm{\hat\phi}_2$.
\end{proposition}

\begin{proof}
Suppose $\norm{w_P}_2>\norm{\hat\phi}_2$. Evaluating \eqref{eq:pen} at $\phi=\hat\phi$ gives
$U(w_P)\ge U(\hat\phi)+c\bigl(\norm{w_P}_2-\norm{\hat\phi}_2\bigr)>U(\hat\phi)$, which contradicts
the optimality of $\hat\phi$.
\end{proof}

So in every window in which the cap does not bind, and the benchmark is then the unconstrained
optimum $\hat\phi$ itself, the benchmark is also the \emph{wider} of the two portfolios. This is a
checkable prediction, and Section~\ref{sec:results} reports that it holds in every slack window of the
sample.

\begin{proposition}[Which slack event, and how often]
\label{prop:half}
Let $X_t=\norm{w_{P,t}}_2$ and $h_t=\frac{1}{t-1}\sum_{s<t}X_s$ for $t\ge2$, and distinguish the
\emph{method-side} event $A_t=\{X_t<h_t\}$ from the \emph{benchmark-side} event
$B_t(\tau)=\{\norm{w_{B,t}}_2<h_t-\tau\}$, $\tau\ge0$. Then:
\begin{enumerate}
\item\label{it:sym} If $(X_1,\dots,X_t)$ is jointly symmetric about $\theta\mathbf{1}$, then
$\Prob(A_t)=\tfrac12\bigl[1-\Prob(X_t=h_t)\bigr]$. In particular $\Prob(A_t)=\tfrac12$ when the
$X_s$ are i.i.d., marginally symmetric and non-atomic. Independence and symmetry are sufficient
here, not necessary.
\item\label{it:stat} If $\{X_t\}$ is strictly stationary and ergodic with $\E|X_t|<\infty$ and no
atom at $m=\E X_t$, then $\Prob(A_t)\to\Prob(X_t<m)$. In the stationary case the limit is one half
only if the mean sits at the median of the marginal law; a skewed breadth distribution moves it.
\item\label{it:bench} If the cap is feasible, the capped program is solved exactly, and the
unconstrained optimum $w_{\mathrm{MV},t}$ of $U_t$ over $\simplex$ is unique, then
$B_t(\tau)\subseteq\{X_t<h_t-\tau\}$. The benchmark slack rate is therefore a different object from
the method-side rate of \ref{it:sym}--\ref{it:stat}, and one does not validate the other.
\end{enumerate}
\end{proposition}

\begin{proof}
\ref{it:sym}: the joint reflection $X_s\mapsto2\theta-X_s$ changes the sign of $X_t-h_t$ and
preserves the law, so the probabilities of $X_t>h_t$ and $X_t<h_t$ coincide and what remains is the
tie mass.

\ref{it:stat}: by the ergodic theorem $h_t\to m$ almost surely; for $\eta>0$ the symmetric
difference of $A_t$ and $\{X_t<m\}$ has probability at most $\Prob(|h_t-m|>\eta)+\Prob(|X_t-m|\le\eta)$,
and letting $t\to\infty$ and then $\eta\downarrow0$ gives the limit.

\ref{it:bench}: if $R(w_{B,t})<h_t$ and $w_{B,t}$ is not the unconstrained optimum, moving a small
amount along the segment towards $w_{\mathrm{MV},t}$ stays inside the cap because $R$ is convex and
improves $U$ because $U$ is concave, contradicting optimality of $w_{B,t}$; hence $w_{B,t}=w_{\mathrm{MV},t}$
whenever the cap is strictly slack. Penalised optimality then gives
$0\le U_t(w_{\mathrm{MV},t})-U_t(w_{P,t})\le c\bigl(R(w_{\mathrm{MV},t})-X_t\bigr)$, so
$R(w_{\mathrm{MV},t})\ge X_t$, which is the claimed inclusion.
\end{proof}

\begin{remark}[What the data can and cannot say about Proposition~\ref{prop:half}]
\label{rem:halfdata}
Three warnings apply here, because this is where a tempting over-claim sits. First, the events in
\ref{it:sym} and \ref{it:bench} are not the same, and Table~\ref{tab:rules} reports the second. In our
data the \emph{method-side} rate is $\Prob(A_t)=60.8\%$ at $\delta=10^{-6}$ and $65.9\%$ at
$\delta=10^{-4}$ over the $812$ panel-windows, against $46\%$ and $10\%$ for $B_t(0.01)$. The two
differ by a factor of two at the tight radius. Second, $60.8\%$ is not a confirmation of
$\tfrac12$ but a measurement of skewness under \ref{it:stat}. The median of $(h_t-X_t)$ is $+0.36$
standard deviations of the panel's own breadth at $\delta=10^{-6}$ and $+0.48$ at $\delta=10^{-4}$, so
the cap sits above the typical method breadth and the width distribution is right-skewed, which is
exactly the case in which the stationary limit $\Prob(X_t<\E X_t)$ exceeds one half. Third, the
windows overlap by about $90\%$, so the effective number of independent observations behind these
percentages is an order of magnitude smaller than $812$. They are descriptive, and we do not attach
a standard error to them.
\end{remark}

\begin{table}[t]
\centering
\small
\caption{Eight cap rules on six panels, both radii. ``Slack'': $\norm{w_B}_2<h-0.01$;
``above'': $\norm{w_B}_2>h+0.01$. The method's own breadth $\norm{w_{P,t}}_2$ is dimensionless with
median $0.48$ at $\delta=10^{-6}$ and $0.34$ at $\delta=10^{-4}$, so the fixed caps at
$N_{\text{eff}}$ effective positions, with $h=N_{\text{eff}}^{-1/2}$, sit far below it, while the
tracking caps sit close to it. $d$ in basis points per day. The oracle rule sets
$h=\norm{w_{P,t}}_2$ in the current window.}
\label{tab:rules}
\begin{tabular}{lcccccc}
\toprule
& \multicolumn{3}{c}{$\delta=10^{-6}$} & \multicolumn{3}{c}{$\delta=10^{-4}$} \\
\cmidrule(lr){2-4}\cmidrule(lr){5-7}
Cap rule & slack & $\overline{d}$ & $\mathrm{sd}(d)$ & slack & $\overline{d}$ & $\mathrm{sd}(d)$ \\
\midrule
fixed, $N_{\text{eff}}=20$ & $0.0\%$ & $+25.1$ & $159$ & $0.0\%$ & $+14.3$ & $103$ \\
fixed, $N_{\text{eff}}=34$ & $0.0\%$ & $+29.5$ & $182$ & $0.0\%$ & $+18.7$ & $124$ \\
fixed, $N_{\text{eff}}=50$ & $0.0\%$ & $+32.5$ & $196$ & $0.0\%$ & $+21.6$ & $138$ \\
mean of past breadth   & $\mathbf{51.4\%}$ & $+8.3$ & $67$ & $13.2\%$ & $+8.6$ & $77$ \\
EWMA of past breadth   & $45.1\%$ & $+7.6$ & $60$ & $9.8\%$ & $+7.0$ & $63$ \\
last past breadth      & $34.8\%$ & $+7.4$ & $50$ & $5.2\%$ & $+7.0$ & $48$ \\
25th pctile of past breadth & $20.1\%$ & $+15.5$ & $100$ & $0.9\%$ & $+16.0$ & $103$ \\
\midrule
current breadth (oracle) & $0.0\%$ & $-0.0$ & $\mathbf{2.1}$ & $0.0\%$ & $0.0$ & $\mathbf{0.0}$ \\
\bottomrule
\end{tabular}
\end{table}

\begin{figure}[t]
\centering
\includegraphics[width=\textwidth]{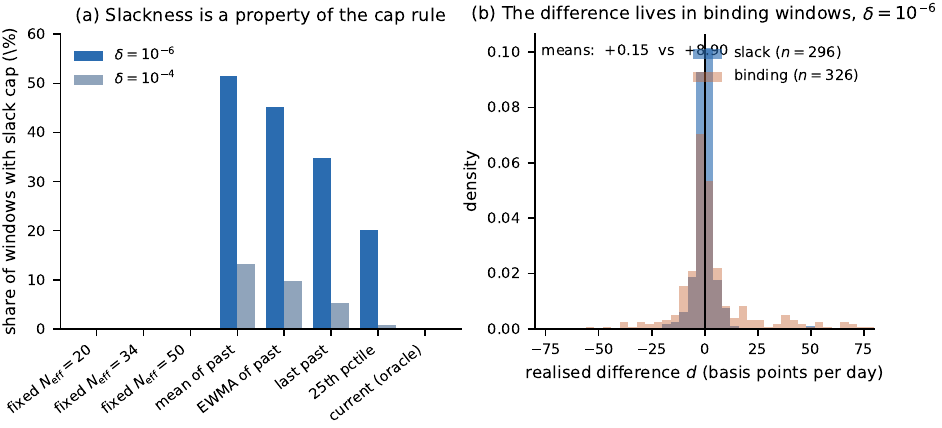}
\caption{(a) Share of windows in which the matched cap is slack, that is, the \emph{benchmark-side}
event $\norm{w_B}_2<h-0.01$, by cap rule and radius, on six panels. We deliberately do not draw the
one-half prediction of Proposition~\ref{prop:half} here. That prediction concerns the
\emph{method-side} event $\{X_t<h_t\}$, which occurs in $61\%$ of windows in this sample, and the two
events are linked only by the inclusion in Proposition~\ref{prop:half}(\ref{it:bench}). (b)
Distribution of the realised difference in slack and binding windows, de-duplicated sample,
$\delta=10^{-6}$. The two regimes are centred at $+0.11$ and $+8.91$ basis points per day
respectively and differ mainly in dispersion.}
\label{fig:rules}
\end{figure}

Table~\ref{tab:rules} and Figure~\ref{fig:rules}(a) report the \emph{benchmark-side} rate
$B_t(0.01)$ for eight cap rules. Proposition~\ref{prop:half} constrains the pattern rather than the
level, and the pattern is there. The tracking rules, whose cap is anchored at the method's own recent
breadth, leave the cap slack in $20\%$--$51\%$ of windows, whereas the fixed caps never do. The oracle
rule, whose cap equals the current realised breadth, reproduces the identity to $2.1$ basis points per
day of standard deviation at $\delta=10^{-6}$ and to zero at $\delta=10^{-4}$.

The reason the fixed caps never slack is not that their medians look small, but an implication of
Proposition~\ref{prop:half}(\ref{it:bench}) that can be checked window by window. Part
\ref{it:bench} says a benchmark can only be slack where the method's own breadth falls below the cap
by more than the tolerance, so if $\norm{w_{P,t}}_2\ge h$ in every window, slackness is impossible. At
$\delta=10^{-6}$ that is what happens. With $N_{\text{eff}}=20,34,50$ and $h=N_{\text{eff}}^{-1/2}$,
the share of panel-windows with $\norm{w_{P,t}}_2\ge h$ is $99.8\%$, $100.0\%$ and $100.0\%$, and the
$0.2\%$ remainder never exceeds the tolerance. At $\delta=10^{-4}$ the argument is weaker. The share is
$65.5\%$, $96.1\%$ and $99.1\%$, so the zero slack rates there are empirical rather than implied. For
the tracking rule the same share is $39.2\%$ and $34.1\%$, which means the method's breadth sits below
its own past mean in the complementary $60.8\%$ and $65.9\%$ of windows. That is the quantity of
Proposition~\ref{prop:half}(\ref{it:stat}) and not the benchmark slack rate.

Scale puts these numbers in perspective. Measured in units of the panel's own breadth standard
deviation, the median gap $h_t-\norm{w_{P,t}}_2$ is $+0.36$ at $\delta=10^{-6}$ and $+0.48$ at
$\delta=10^{-4}$, the median gap $h_t-\norm{w_{B,t}}_2$ is $+0.04$ and $+0.00$, and the $0.01$
tolerance is $0.08$ in the same units. The cap therefore sits well above the typical method breadth
while the benchmark solution sits essentially on it. What separates the $61\%$ method-side rate from
the $46\%$ benchmark-side rate is the distance from the method's breadth to the cap and not the
tolerance.

\subsection{Why there is no alpha to test for this class of methods}

It is worth being blunt about what $d_t$ can detect. Because $w_P$ solves \eqref{eq:dro},
Proposition~\ref{prop:collapse} says it is also the maximiser of $U$ over
$\{\phi\in\simplex:\norm{\phi}_2\le\norm{w_P}_2\}$. Hence a comparison of $w_P$ against an
\emph{ex ante} optimum at the \emph{same} breadth is vacuous, and that optimum is the object a
breadth-matched study is meant to construct. Formally, define the shortfall of $w_P$ relative to the
frontier at its own breadth,
\begin{equation}
A_t \;=\; \max_{\phi\in\simplex,\ \norm{\phi}_2\le\norm{w_{P,t}}_2}
\bigl[\phi^\top\hat\mu_t-\lambda\phi^\top\hat\Sigma_t\phi\bigr]
\;-\;\bigl(w_{P,t}^\top\hat\mu_t-\lambda w_{P,t}^\top\hat\Sigma_t w_{P,t}\bigr).
\label{eq:A}
\end{equation}
Then $A_t\equiv0$ \emph{by definition}, because the feasible set in \eqref{eq:A} is the one that
Proposition~\ref{prop:identity} characterises, and this holds for every method of the form
\eqref{eq:dro} and for any $\hat\Sigma\succeq0$. Note what the statement does and does not say. It
concerns the \emph{estimated} objective, with $\hat\mu_t$ and $\hat\Sigma_t$ held fixed, so $w_P$ is
optimal for the in-sample problem at its own breadth and not necessarily ex post or in population. A
test of whether the excess survives breadth matching therefore has nothing to test in this class,
since the quantity is zero for any inputs. This is stronger than \emph{unidentified}. It is
identically zero, and that is why the present paper is an audit rather than an alpha test. Testing
alpha requires a method whose information is not already contained in $(\hat\mu,\hat\Sigma)$, and we
do not have one here.

A related point about the identity concerns direction. Proposition~\ref{prop:identity} gives
$g=0\Rightarrow w_B=w_P\Rightarrow d=0$, so agreement of the two breadth measures is a
\emph{sufficient} structural condition for a zero difference. The converse does not hold in general,
since even with $g\neq0$ the two portfolios can earn the same realised return in a given window. A zero
observed difference is therefore not evidence that $g=0$, and the model restrictions in
Section~\ref{sec:results} are of the form $m(0,x)\equiv\E[d\mid g=0,x]=0$ rather than of the form
$d=0$ if and only if $g=0$.

\subsection{Two mismatch variables, and why the distinction matters}

The two gaps of \eqref{eq:g}--\eqref{eq:geff} coincide only when the cap binds. When the cap is
slack, $w_{B,t}$ is the unconstrained mean--variance portfolio and $\norm{w_{B,t}}_2<h_t$, so
$g_t$ overstates the effective gap. By Proposition~\ref{prop:shrink} the actual gap is then
\emph{negative}. Both gaps are predetermined, since $\norm{w_{P,t}}_2$, $h_t$ and $w_{B,t}$ are all
known when window $t$ begins, whereas $d_t$ is realised afterwards.

\begin{remark}[$g^{\text{eff}}$ also satisfies the restriction in slack windows]
\label{rem:geffslack}
If $g^{\text{eff}}_t=0$ and the cap is slack, then $\norm{w_{B,t}}_2=\norm{w_{P,t}}_2<h_t$. Since
$w_{B,t}$ maximises $U$ over the larger ball $\{\norm{\phi}_2\le h_t\}$ it also maximises $U$ over
the smaller bounded set $\{\phi\in\simplex:\norm{\phi}_2\le\norm{w_{P,t}}_2\}$, which by
Proposition~\ref{prop:identity} has $w_{P,t}$ as its unique maximiser; hence $w_{B,t}=w_{P,t}$ and
$d_t=0$. So restriction \eqref{eq:restrict} holds for $g^{\text{eff}}$ in the slack regime as well
as in the binding one, which is what makes it usable as an estimand at all.
\end{remark}

\subsection{The cap is often not binding}

Table~\ref{tab:slack} reports the distribution of $\norm{w_{B,t}}_2-h_t$. At $\delta=10^{-6}$ the
absolute deviation exceeds $0.01$ in $47.3\%$ of unique calendar windows and $0.05$ in $30.4\%$,
while at $\delta=10^{-4}$ the cap binds much more tightly. Decomposing by sign matters here. At
$\delta=10^{-6}$, $53.8\%$ of panel-windows are within $0.01$ of the target and $46.2\%$ fall
\emph{below} it, which is genuine slackness, since the unconstrained optimum already satisfies the
cap. None exceed it at the $1500$-iteration tolerance. The same pattern holds at $\delta=10^{-4}$,
with $89.7\%$, $10.3\%$ and $0.0\%$. The large deviations are therefore slackness and not solver
failure.

The claim that the benchmark is then \emph{the} mean--variance portfolio is verifiable directly.
In the slack windows the $\ell_1$ distance between $w_B$ and the unconstrained mean--variance
solution has median $1.5\times10^{-3}$ at $\delta=10^{-6}$ ($p_{90}=0.044$) and
$1.4\times10^{-3}$ at $\delta=10^{-4}$; in the binding windows the same distance has median
$0.263$ and $0.692$. The two populations are cleanly separated, so in slack windows the
``breadth-matched'' benchmark is the unconstrained mean--variance portfolio to three decimal
places.

This is an audit finding in its own right, and an easily avoidable one. A study that reports a
``breadth-matched'' comparison should report the slack distribution, because in the slack windows the
benchmark is not a breadth-matched portfolio at all. It also means that specifications written in
terms of the target gap $g_t$ describe the \emph{rule} and not the realised breadth difference. We
report both and take $g_t^{\text{eff}}$ as the estimand of record, as noted in
Remark~\ref{rem:geffslack}. The bisection solves for the ridge multiplier on a log grid over
$[10^{-10},10^{4}]$ in $18$ steps with $220$ projected-gradient iterations per step.

\paragraph{Slackness is not solver error.} Four checks, because a benchmark that is off its cap
could be slack for real or could be badly solved. First, and most directly, Table~\ref{tab:slack}
and every table that follows are computed at $1500$ inner iterations per bisection step rather than
the default $220$; at that tolerance \emph{no} window exceeds its cap, against $0.4\%$ before, while
the slack share moves from $47.6\%$ to $47.3\%$ and every headline number is unchanged to the second
decimal (the binding-window mean difference is $+8.91$ against $+8.90$, and the raw slope is $63.25$
against $63.20$). The quantity that moves with the solver's accuracy is the violation residue. The
quantity that does not is slackness. Second, an independent budget scan on six panels, using $220$,
$1000$ and $3000$ gradient steps, leaves the slack share \emph{exactly} unchanged at $32.3\%$ while
the residual above-cap share falls from $0.29\%$ to $0.00\%$. Third, the oracle rule provides a
direct precision test, since its cap is by construction the method's own breadth. At $220$ inner
iterations the benchmark exceeds its cap in $6.3\%$ of windows by up to $4.5\times10^{-2}$; at $1500$
iterations the exceedance share is $0.0\%$ with a maximum of $0$, and the ninety-percentile absolute deviation
falls from $6.3\times10^{-3}$ to $1.2\times10^{-6}$. The reported violation rates are therefore
solver residues that vanish under a tighter tolerance, whereas the reported slack rates are
invariant to it. Third, in slack windows the benchmark is the unconstrained mean--variance portfolio
not merely by construction but observably, and Proposition~\ref{prop:shrink} says it must then be the
wider of the two portfolios. The inequality $\norm{w_P}_2\le\norm{w_B}_2$ holds in $375/375$ slack
panel-windows at $\delta=10^{-6}$ and $84/84$ at $\delta=10^{-4}$, without exception.

\begin{table}[t]
\centering
\small
\caption{Mismatch and cap slackness, 14 panels $\times$ 58 windows, de-duplicated into unique
calendar windows ($n=622$), all quantities at $1500$ inner iterations. $g$: target gap \eqref{eq:g}.
Slack: $\abs{\norm{w_B}_2-h_t}$, split by sign, with the tolerance $0.01$. No window exceeds its cap
at this tolerance, so the ``above'' column is zero by construction rather than by rounding.}
\label{tab:slack}
\begin{tabular}{lcccc}
\toprule
$\delta$ & median $g$ & share $\abs{g}>0.05$ & share slack $>0.01$ & share above cap $>0.01$ \\
\midrule
$10^{-6}$ & $-0.025$ & 64.6\% & 47.3\% & 0.0\% \\
$10^{-4}$ & $-0.037$ & 66.2\% & 10.8\% & 0.0\% \\
\bottomrule
\end{tabular}
\end{table}

Consistent with this, the realised difference is far larger than the solver residue of
Table~\ref{tab:conv}. Taking the mean over each 10-day holding period and then over windows, the
per-panel mean difference has median $+5.7$ basis points per day, with range $-2.3$ to $+10.2$, and
the per-panel dispersion of the window means has median $26.9$, with range $7.0$ to $40.7$, at
$\delta=10^{-6}$. At $\delta=10^{-4}$ the corresponding figures are $+4.9$, with range $-0.7$ to
$+10.0$, and $27.7$, with range $5.1$ to $35.9$. These are dispersions of holding-period averages and
not of daily returns, and the distinction matters. For comparison, the self-referential construction
gives $0.03$--$0.43$ on the same scale. On the six-panel archival run with five alternative tracking
rules the daily figures are $48$--$79$ bp/day of dispersion and $+7$ to $+11$ bp/day of mean.

\section{Data and protocol}
\label{sec:data}

We use 14 panels of daily returns on $N=255$ Chinese A-share stocks, each covering 648 trading
days, with start dates staggered by about one calendar quarter from March 2018 to June 2021. Each
panel spans about 31 months while starts are about three months apart, so consecutive panels overlap
by roughly $90\%$, which means the panels are \emph{not} 14 independent replications. We handle this
by de-duplicating panel-windows into unique calendar windows, so that 812 panel-windows become 622
unique calendar windows, and by clustering standard errors on blocks of 20 consecutive unique windows.

We use a 60-day estimation window and a 10-day holding period, giving 58 non-overlapping
out-of-sample windows and 580 out-of-sample days per panel. Covariance estimates use Ledoit--Wolf
shrinkage for $\hat\Sigma$ \citep{ledoit2004}, the risk aversion parameter is $\lambda=10$, and the
radii are $\delta\in\{10^{-6},10^{-4}\}$. The method is \eqref{eq:dro} and the benchmark is
\eqref{eq:track}, with $\mathcal{R}$ the mean of past realised breadths. In the first window no
history exists, so we use the reference breadth $N_{\text{eff}}=34$, that is
$h_1=34^{-1/2}\approx0.171$, which affects one window out of 58 per panel and is included throughout.
The value is inherited from the ex-ante cap grid used to select the reference breadth in
\citet{chen2026breadth}.

Both programs are solved by projected gradient ascent on the simplex with step $1/L$, $L$ the
Lipschitz constant of the gradient, 400 iterations for the empirical work, and up to 3000 for the
identity diagnostics. The capped program \eqref{eq:track} is solved by bisection on the ridge
multiplier so that the realised norm hits $h_t$; \S\ref{sec:results} reports how well that
succeeds.

\section{Results}
\label{sec:results}

\subsection{The identity-consistent specification}

The identity implies a conditional-mean restriction and nothing more:
\begin{equation}
m(g,x)\;=\;\E[d\mid g=g_0,\,x]\;=\;0\qquad\text{at } g_0=0,\ \text{for every } x .
\label{eq:restrict}
\end{equation}
The general form satisfying \eqref{eq:restrict} is
\begin{equation}
d \;=\; g\,\phi(x)+\varepsilon,
\label{eq:phi}
\end{equation}
with $\phi$ an arbitrary function; a linear interaction, $\phi(x)=\alpha+\beta x$, is only the
lowest-dimensional member of that class. This distinction matters. A specification with an
intercept and a separate $\norm{w_P}_2$ term does not satisfy \eqref{eq:restrict}: whatever it
estimates at $g=0$ is unrestricted and in general non-zero, so it cannot test the restriction. We
report such a specification in Appendix~\ref{app:naive} only as a description of in-sample patterns,
and the fact that its median stratum intercept is not far from zero in our sample is not evidence
for the restriction, because it was free to be anything.

\paragraph{Declaring the estimand.} The two mismatch measures \eqref{eq:g}--\eqref{eq:geff} answer
different questions. The quantity $g$ is what the rule \emph{targets} and $g^{\text{eff}}$ is the
breadth gap the benchmark \emph{achieves}. Because this paper is about breadth, we take the achieved
gap $g^{\text{eff}}$ as the estimand of record and report $g$ as the rule-level counterpart. The two
are not interchangeable and we do not pool them.

We estimate $d=g\,\phi(x)+\varepsilon$ with no independent intercept, taking $\phi$ piecewise linear
with knots fixed in advance at the quartiles of $x=\norm{w_P}_2$, which gives three knots and five
coefficients. Standard errors are clustered on blocks of 20 consecutive unique windows, that is 32
blocks. Because that is few, we report wild cluster bootstrap $p$-values, using Rademacher weights and
4999 replications with the null imposed by the restricted residuals, alongside the asymptotic ones.

Table~\ref{tab:phi} reports the outcome, and all $p$-values in it come from a single run with $4999$
bootstrap replications, whose Monte Carlo standard error at the $5\%$ level is $0.2$ percentage
points. Reading the estimand of record first, under $g^{\text{eff}}$ the no-effect null has
$\mathrm{WCR}\;p=0.048$ at $\delta=10^{-6}$ and $0.048$ at $\delta=10^{-4}$. Under the rule-level gap
$g$ it is $0.052$ and $0.050$. Against the asymptotic Wald $p<0.001$ in all four cells this is a large
difference, and with $32$ clusters we do not rely on the asymptotic number.

The point estimates have a shape that is stable across all four cells. The return to mismatch is
around zero or negative at the tenth and fiftieth percentiles of breadth and large and positive at the
ninetieth. Under $g$ at $\delta=10^{-6}$ the estimates are $-5.0$, $-18.2$ and $+120.7$ at
$x=0.32$, $0.39$ and $0.66$. The other three cells give $-26.4$, $+3.2$, $+123.7$; $-7.6$, $-2.2$,
$+128.8$; and $-5.6$, $-3.2$, $+132.2$. What we cannot do is attach a shape \emph{test}. The
linearity null is rejected under $g$ at $\delta=10^{-6}$ ($\mathrm{WCR}\;p=0.019$) but not in the
other three cells ($0.347$, $0.593$, $0.574$), and a three-parameter shape test with $32$ clusters
has little power here. We therefore describe the shape and do not test it. We also do not read the
three non-rejections as evidence of linearity, since the point estimates in those very cells are
plainly non-linear.

Note the disagreement between the asymptotic and bootstrap linearity tests. Under $g^{\text{eff}}$
at $\delta=10^{-6}$ the asymptotic $p$ is $0.040$ while the bootstrap $p$ is $0.347$. With $32$
clusters the asymptotic test rejects far too readily, which is precisely why every number in this
subsection is a wild cluster bootstrap number.

\begin{table}[t]
\centering
\small
\caption{The identity-consistent model $d=g\,\phi(\norm{w_P}_2)+\varepsilon$, no independent
intercept, $\phi$ piecewise linear with knots at the quartiles of $\norm{w_P}_2$. De-duplicated
sample, $n=622$; $4999$ bootstrap replications. The no-effect column tests $\phi\equiv0$; the
linearity column tests that the three knot coefficients vanish, i.e.\
$\phi(x)=\alpha+\beta x$. Wald: asymptotic $\chi^2$; WCR: wild cluster bootstrap with $32$ blocks.}
\label{tab:phi}
\begin{tabular}{lcccccc}
\toprule
& & & \multicolumn{2}{c}{$H_0:\phi\equiv0$} & \multicolumn{2}{c}{$H_0:\phi$ linear} \\
\cmidrule(lr){4-5}\cmidrule(lr){6-7}
$\delta$ & mismatch & $\hat\phi$ at $p_{10}$, $p_{50}$, $p_{90}$ & Wald $p$ & WCR $p$ & Wald $p$ & WCR $p$ \\
\midrule
$10^{-6}$ & $g$             & $-5,\ -18,\ +121$ & $<0.001$ & $0.052$ & $<0.001$ & $\mathbf{0.019}$ \\
$10^{-6}$ & $g^{\text{eff}}$ & $-27,\ +3,\ +124$ & $<0.001$ & $0.048$ & $0.040$ & $0.347$ \\
$10^{-4}$ & $g$             & $-7,\ -2,\ +129$ & $<0.001$ & $0.050$ & $0.343$ & $0.593$ \\
$10^{-4}$ & $g^{\text{eff}}$ & $-5,\ -3,\ +132$ & $<0.001$ & $0.048$ & $0.304$ & $0.574$ \\
\bottomrule
\end{tabular}
\end{table}

\subsection{Auxiliary description, and why it is not the identification}

A natural summary is a no-intercept stratification,
\begin{equation}
d \;=\; g\sum_{k=1}^{4}\gamma_k\mathbf{1}\{x\in\text{quartile }k\}+\varepsilon ,
\label{eq:gamma}
\end{equation}
which does respect \eqref{eq:restrict}, unlike a conventional stratified regression. At
$\delta=10^{-6}$ the estimates are $(\gamma_1,\dots,\gamma_4)=(-4,-11,15,124)$ with stratum means
$\bar x=(0.32,0.37,0.43,0.62)$. Equality of the four is rejected
($\mathrm{WCR}\;p=0.001$), the estimates are not monotone since the second is below the first, and a
linear fit in $\bar x$ gives slope $+471$. At $\delta=10^{-4}$,
$(\gamma_1,\dots,\gamma_4)=(-13,13,15,124)$, equality is rejected ($\mathrm{WCR}\;p=0.004$), and here
the estimates \emph{are} monotone. These figures come from the same $4999$-replication run as
Table~\ref{tab:phi}, and they agree with it. The top stratum, at a mean breadth of $0.62$, carries a
return to mismatch of $124$ bp/unit, which is the same order as the $\hat\phi$ of $121$--$132$
evaluated at the ninetieth percentile of breadth in that table.

Two cautions attach to \eqref{eq:gamma}. First, the conventional stratified regression with
within-stratum intercepts does \emph{not} respect \eqref{eq:restrict}: its stratum-specific
intercepts allow $d\neq0$ at $g=0$. Its significance cannot be used to overturn a well-specified
interaction, and it corresponds to a different estimand; we report it only in
Appendix~\ref{app:naive} and label it descriptive. Second, non-monotonicity at
$\delta=10^{-6}$ means that ``the return rises with concentration'' is too crude a summary even
here.

\subsection{What the difference is made of: a return premium for risk, not a value gap}

Proposition~\ref{prop:return} is an identity, and every term on its right-hand side is computable
from the pipeline except $m_t$ and $\varepsilon_t$. We therefore report the decomposition directly
rather than inferring it from a slope. Table~\ref{tab:regime} splits the de-duplicated sample by
whether the cap is at its target and reports three means. They are the raw difference $d$, the
\emph{risk-adjusted} difference $d-\lambda\Delta\mathrm{Risk}$, and the in-sample value gap
$c\,g+q$, which equals $U_t(w_P)-U_t(w_B)$ by Proposition~\ref{prop:return}.

\begin{table}[t]
\centering
\small
\caption{Decomposition of the realised difference, de-duplicated sample ($n=622$), basis points per
day, with the inner solver run to $1500$ iterations per bisection step (no window violates its cap
at that tolerance, against $0.4\%$ at the default budget). Each mean column carries its clustered
$t$-statistic in parentheses for the two return measures (standard errors clustered on blocks of 20
consecutive windows). ``Slack'': the cap is not at its target, $\abs{\norm{w_B}_2-h_t}>0.01$. The
risk adjustment uses the coefficient of risk aversion the method was run with, $\lambda=10$. The
value gap $\overline{c\,g}+\bar q$ equals $\overline{U_t(w_P)-U_t(w_B)}$ and is the object the
identity constrains; the residual $\overline{m+\varepsilon}$ closes the identity exactly, and the
rows are weighted so that the pooled row is the weighted average of the two regime rows.}
\label{tab:regime}
\begin{tabular}{lccccccc}
\toprule
$\delta$ & window & $n$ & $\overline{d}$ & $\overline{d-\lambda\Delta\mathrm{Risk}}$ &
$\overline{c\,g}$ & $\bar q$ & $\overline{m+\varepsilon}$ \\
\midrule
\multirow{2}{*}{$10^{-6}$} & slack   & 294 & $+0.11\ (0.29)$ & $+0.85\ (2.04)$  & $-1.01$ & $+1.10$ & $+0.76$ \\
                           & binding & 328 & $+8.91\ (2.26)$ & $-4.37\ (-1.74)$ & $+0.76$ & $+5.39$ & $-10.52$ \\
\cmidrule(lr){2-8}
                           & pooled  & 622 & $+4.75\ (2.50)$ & $-1.90\ (-1.18)$ & $-0.08$ & $+3.37$ & $-5.19$ \\
\midrule
\multirow{2}{*}{$10^{-4}$} & slack   & 67  & $+0.36\ (0.09)$ & $+6.80\ (1.73)$  & $-15.75$ & $+10.98$ & $+11.58$ \\
                           & binding & 555 & $+5.66\ (2.40)$ & $+1.39\ (0.86)$  & $+0.64$ & $+4.94$ & $-4.20$ \\
\cmidrule(lr){2-8}
                           & pooled  & 622 & $+5.09\ (2.49)$ & $+1.97\ (1.24)$  & $-1.12$ & $+5.59$ & $-2.50$ \\
\bottomrule
\end{tabular}
\end{table}

Three readings follow, and they are less flattering than the raw numbers.

First, the split by cap status is large in raw return, and it is not a restatement of the fact that
the two portfolios sometimes coincide. In slack windows the raw difference is $+0.11$ bp/day, and there
are $294$ such windows at $\delta=10^{-6}$. In binding windows it is $+8.91$ bp/day, some $2.3$
standard errors from zero, and the split is exhaustive with exact arithmetic, since
$0.473\times0.11+0.527\times8.91=4.75$ reproduces the pooled mean. The obvious objection is that where
the cap is slack the benchmark is the unconstrained optimum and the method may be close to it, so that
a near-zero difference would be mechanical. Table~\ref{tab:weights} rejects that explanation. In slack
windows the benchmark is indeed the unconstrained optimum, with a median $\norm{w_B-w_{\mathrm{MV}}}_1$
of $0.006$, yet the method is \emph{not} close to it. The median $\norm{w_P-w_B}_1$ is $0.114$ at
$\delta=10^{-6}$ and $0.721$ at $\delta=10^{-4}$, against $0.207$ and $0.435$ in binding windows. Two
clearly different portfolios therefore earn indistinguishable returns in slack windows, and whatever
makes the difference in binding windows is not the mere distance between the two weight vectors.

\begin{table}[t]
\centering
\small
\caption{Weight distances by regime, median $\ell_1$ distance, de-duplicated sample.
$w_{\mathrm{MV}}$ is the unconstrained mean--variance optimum at the same estimates. In slack windows
the benchmark is the unconstrained optimum to three decimal places, so the near-zero realised
difference cannot be explained by the method being close to the benchmark.}
\label{tab:weights}
\begin{tabular}{llcccc}
\toprule
$\delta$ & window & $n$ & $\norm{w_P-w_B}_1$ & $\norm{w_P-w_{\mathrm{MV}}}_1$ & $\norm{w_B-w_{\mathrm{MV}}}_1$ \\
\midrule
$10^{-6}$ & slack   & 294 & $0.114$ & $0.104$ & $0.006$ \\
$10^{-6}$ & binding & 328 & $0.207$ & $0.080$ & $0.284$ \\
$10^{-4}$ & slack   & 67  & $0.721$ & $0.752$ & $0.017$ \\
$10^{-4}$ & binding & 555 & $0.435$ & $0.801$ & $0.714$ \\
\bottomrule
\end{tabular}
\end{table}

Second, \emph{the pooled raw difference does not survive the risk adjustment, and neither does its
sign}. At $\lambda=10$ the adjusted difference $d-\lambda\Delta\mathrm{Risk}$ averages $-1.90$ bp/day
with a clustered standard error of $1.61$, which is indistinguishable from zero, and in binding
windows it averages $-4.37$ with a $t$-statistic of $-1.74$, also not significant at $5\%$. The risk
difference is worth $+\lambda\overline{\Delta\mathrm{Risk}}=+6.64$ bp/day in the same objective
function the method maximises.

The point estimate $-1.90$ must not be read as an established sign, because the adjustment is not
estimated from the data. It is the method's own $\lambda$, and $\lambda$ is a tuning parameter of the
technology under test rather than a measured preference of any investor. Table~\ref{tab:lambda} scans
it. The adjusted mean is linear in $\lambda$ and crosses zero at $\lambda^{*}=7.14$ at
$\delta=10^{-6}$ and $\lambda^{*}=16.33$ at $\delta=10^{-4}$. At $\lambda=5$ the tight-radius figure is
$+1.42$, and at $\lambda=15$ it is $-5.22$. The honest statement is therefore narrower than a claim
that the excess is a risk premium. At the radius and risk aversion we ran, the risk-adjusted
difference \emph{cannot be distinguished from zero at either radius}, whereas the raw difference can be
distinguished from zero at both. The raw excess is therefore not established to be anything other than
compensation for risk taken, and whether it is exactly that depends on a parameter we do not observe.

\begin{table}[t]
\centering
\small
\caption{The risk adjustment under alternative coefficients of risk aversion. $\lambda$ is the
method's own tuning parameter ($\lambda=10$ in all other tables). Entries are means of
$d-(\lambda/10)\lambda\Delta\mathrm{Risk}$ in basis points per day, with clustered $t$-statistics in
parentheses; $\lambda^{*}$ is the value at which the mean crosses zero. This is a partial-equilibrium
scan. Only the multiplier changes, not the portfolios, so it answers how risk-averse an investor would
have to be for the measured excess to disappear, and not what would happen if the method were run at
another $\lambda$.}
\label{tab:lambda}
\begin{tabular}{lcccccc}
\toprule
$\delta$ & $\lambda^{*}$ & $\lambda=2$ & $\lambda=5$ & $\lambda=10$ & $\lambda=15$ & $\lambda=20$ \\
\midrule
$10^{-6}$ & $7.14$ & $+3.42\ (2.14)$ & $+1.42\ (1.07)$ & $-1.90\ (-1.18)$ & $-5.22\ (-2.12)$ & $-8.55\ (-2.45)$ \\
$10^{-4}$ & $16.33$ & $+4.47\ (2.49)$ & $+3.53\ (2.30)$ & $+1.97\ \ (1.24)$ & $+0.41\ \ (0.19)$ & $-1.14\ (-0.38)$ \\
\bottomrule
\end{tabular}
\end{table}

Third, the in-sample value gap $c\,g+q=U_t(w_P)-U_t(w_B)=V_t(p_t)-V_t(h_t)$ is exactly what the
identity pins down, yet its \emph{sign} is not a prediction of the theory. Since $V$ is
non-decreasing, the gap has the sign of $g_t$, positive when the method is wider than the cap and
negative or zero when the cap is slack. The pooled figure of $+3.29$ bp/day, with a standard error of
$0.98$, therefore means only that most windows bind and carry $g>0$. It is not a quantity the identity
forces to be positive. The regime rows make the point concretely. In slack windows, where
Proposition~\ref{prop:shrink} forces $g<0$, the theory permits only values $\le0$. At
$\delta=10^{-4}$ the estimate is $-4.77$ with a $t$-statistic of $-9.00$, exactly as it should be,
while at $\delta=10^{-6}$ it is $+0.09$ with a standard error of $0.98$, which we read as the residue
of solving the capped program to a finite tolerance rather than as evidence either way. The sign of
this gap is not something the data can decide against the theory, because the theory fixes it up to
the sign of $g$.

What does not carry over to realised returns is a separate matter, and there the numbers do bite.
The residual $m+\varepsilon$ has unconditional mean $\E[m_t]$, since the innovation term has zero
conditional mean. It averages $-5.19$ bp/day at $\delta=10^{-6}$ with a clustered standard error of
$2.31$ with a $t$-statistic of $-2.24$, and $-2.49$ with a standard error of $3.30$ at
$\delta=10^{-4}$, a $t$-statistic of $-0.75$. The negative sign at the tight radius is not a
discovery. It is the optimisation bias long associated with estimated frontier problems
\citep{michaud1989}, in which a benchmark chosen as the in-sample optimum of estimated inputs carries
an upward-biased in-sample expected value. Appendix~\ref{app:e7} measures that bias once the truth is
known. In that known-truth simulation the decomposition closes to $3.6\times10^{-14}$ bp, $m_t$ is the
largest single term, and $\mathrm{corr}(g,m)=-0.89$. The appendix also shows that the \emph{sign} of
the bias is not a law, since it reverses across the two radii in the same design, so our decomposition
supplies a measurement of the bias in these data rather than a sign theorem. At the tight radius the
measured bias is $-5.2$ bp/day and is large enough to absorb the whole positive in-sample value gap,
while at the wide radius it is not distinguishable from zero. The sample mean of $m+\varepsilon$ is
itself noisy, so we report it with its clustered standard error rather than treating $-5.19$ as a
point estimate of $\E[m_t]$.

\begin{remark}[What the regime split is and is not]
\label{rem:regime}
Membership in the two groups is determined by the same estimates that enter $d_t$, so the entries of
Table~\ref{tab:regime} are conditional descriptions $\E[\cdot\mid S]$, not a causal decomposition;
the $t$-statistics test whether a conditional mean is zero, not whether the cap caused anything.

The asymmetry between the groups is informative all the same, once the two currencies are kept
apart. In \emph{raw return} the method does better in binding windows, where the figure is $+8.91$
bp/day, and not in slack windows, where it is $+0.11$. In \emph{risk-adjusted return} the sign flips,
giving $-4.37$ and $+0.85$. In the \emph{in-sample objective} the value gap $V_t(p_t)-V_t(h_t)$ is
positive in binding windows, at $+6.18$ where $g>0$, and indistinguishable from zero in slack ones, at
$+0.09$ where the theory permits only values $\le0$. These three readings are not alternatives. They
are different currencies, and the theory speaks only to the third. The defensible summary is therefore
narrow. The benchmark does worse on raw return exactly where it is constrained below its own optimum,
and that is a statement about raw returns in this sample rather than the verdict of the theory.
\end{remark}

\subsection{A parsimonious summary, and what its slope is made of}

The identity-consistent model \eqref{eq:phi} nests a constant return to mismatch,
\begin{equation}
d_t \;=\; b\,g_t+\varepsilon_t ,
\label{eq:slope}
\end{equation}
and estimating \eqref{eq:slope} by no-intercept least squares gives the numbers in
Table~\ref{tab:slope}. Two features are worth separating. First, $b$ is not the slope of the return
to mismatch at a given concentration; because \eqref{eq:phi} has $\phi(x)=\alpha+\beta x$, the
constant-slope coefficient estimates $\alpha+\beta\,\E_w[x]$ with weights proportional to $g^2$,
i.e.\ the \emph{mismatch-weighted average} of the return to mismatch. That is a legitimate estimand
and it is reported as such, yet it is not $\phi$ evaluated anywhere in particular, and with
$\mathrm{corr}(g,\norm{w_P}_2)=0.81$--$0.85$ in this sample the two are not interchangeable.
Second, and this is the point of Corollary~\ref{cor:projection}, the slope is not a test of whether
the identity holds. It differs from $c$ by three projection terms, and the honest question is which
of them accounts for the difference.

\begin{table}[t]
\centering
\small
\caption{The constant-slope summary \eqref{eq:slope}, de-duplicated sample ($n=622$), standard
errors clustered on $32$ blocks of 20 consecutive windows. ``Risk-adjusted'' subtracts
$\lambda\Delta\mathrm{Risk}_t$; the third row also subtracts the two other observable terms, $c\,g_t$
and $q_t$, so that its slope is exactly the projection of the mean-forecast error on the mismatch
(matching the last column of Table~\ref{tab:slopedecomp}). $c=\sqrt{\delta}$ in basis points per unit
of $g$ is $10$ at $\delta=10^{-6}$ and $100$ at $\delta=10^{-4}$. WCR: wild cluster bootstrap, 4999
replications throughout, the same count as in Table~\ref{tab:phi}.}
\label{tab:slope}
\begin{tabular}{llcccccc}
\toprule
& & & & \multicolumn{2}{c}{WCR $p$ for} & \\
\cmidrule(lr){5-6}
$\delta$ & dependent variable & mismatch & $\hat b$ & s.e. & $H_0:b=0$ & $H_0:b=c$ & $\hat b/c$ \\
\midrule
$10^{-6}$ & $d$                     & $g$             & $63.25$  & $20.39$ & $0.019$ & $0.029$ & $6.33$ \\
$10^{-6}$ & $d$                     & $g^{\text{eff}}$ & $110.33$ & $26.46$ & $0.009$ & $0.019$ & $11.03$ \\
$10^{-6}$ & $d-\lambda\Delta\mathrm{Risk}$ & $g$        & $-37.03$ & $17.59$ & $0.076$ & $0.024$ & $-3.70$ \\
$10^{-6}$ & $d-\lambda\Delta\mathrm{Risk}$ & $g^{\text{eff}}$ & $-57.20$ & $27.83$ & $0.112$ & $0.067$ & $-5.72$ \\
$10^{-6}$ & $d-cg-q-\lambda\Delta\mathrm{Risk}$ & $g$  & $-89.62$ & $24.11$ & $<0.001$ & $<0.001$ & $-8.96$ \\
$10^{-6}$ & $d-cg-q-\lambda\Delta\mathrm{Risk}$ & $g^{\text{eff}}$ & $-147.16$ & $29.06$ & $0.012$ & $0.006$ & $-14.72$ \\
\midrule
$10^{-4}$ & $d$                     & $g$             & $72.04$  & $20.83$ & $0.016$ & $0.179$ & $0.72$ \\
$10^{-4}$ & $d$                     & $g^{\text{eff}}$ & $80.54$ & $21.09$ & $0.016$ & $0.371$ & $0.81$ \\
$10^{-4}$ & $d-\lambda\Delta\mathrm{Risk}$ & $g$        & $-26.34$ & $17.38$ & $0.146$ & $<0.001$ & $-0.26$ \\
$10^{-4}$ & $d-\lambda\Delta\mathrm{Risk}$ & $g^{\text{eff}}$ & $-25.92$ & $19.12$ & $0.193$ & $<0.001$ & $-0.26$ \\
$10^{-4}$ & $d-cg-q-\lambda\Delta\mathrm{Risk}$ & $g$  & $-149.89$ & $24.05$ & $<0.001$ & $<0.001$ & $-1.50$ \\
\bottomrule
\end{tabular}
\end{table}

\begin{table}[t]
\centering
\small
\caption{Exact decomposition of the estimated slope per Corollary~\ref{cor:projection}. The four
contributions sum to $\hat b$ by construction, and the pipeline asserts this, in basis points per unit
of the mismatch variable. The price contribution is $\E_n[x\,c\,g]/\E_n[x^2]$ evaluated with the
target gap, which equals $c$ exactly when the regressor $x$ is $g$ itself and is slightly larger when
$x$ is $g^{\text{eff}}$.}
\label{tab:slopedecomp}
\begin{tabular}{llcccccc}
\toprule
$\delta$ & mismatch & $c\,g$ & $q_t$ & $\lambda\Delta\mathrm{Risk}_t$ & $m_t+\varepsilon_t$ & sum $=\hat b$ \\
\midrule
$10^{-6}$ & $g$             & $10.00$  & $+42.60$  & $+100.28$ & $-89.62$  & $63.25$ \\
$10^{-6}$ & $g^{\text{eff}}$ & $10.88$ & $+79.08$  & $+167.53$ & $-147.16$ & $110.33$ \\
$10^{-4}$ & $g$             & $100.00$ & $+23.54$  & $+98.38$  & $-149.89$ & $72.04$ \\
$10^{-4}$ & $g^{\text{eff}}$ & $104.00$ & $+31.03$ & $+106.46$ & $-160.96$ & $80.54$ \\
\bottomrule
\end{tabular}
\end{table}

The decomposition makes two things visible that the raw slope conceals.

First, \emph{the risk term dominates}. Table~\ref{tab:slopedecomp} shows that at $\delta=10^{-6}$
the risk term contributes $+100$ to a slope of $63$. On its own it is larger than the coefficient it
is meant to explain, and it is offset by $-90$ from the mean-forecast and innovation terms. A study
that reports only the raw return difference at matched breadth is therefore, in this construction,
reporting mostly the difference in risk taken, priced at the same $\lambda$ that the method
maximises. This is the caveat of Remark~\ref{rem:notidentity}, quantified.

Second, \emph{the identity does not imply $\hat b=c$, and the data do not deliver it}. Removing
$\lambda\Delta\mathrm{Risk}_t$ gives $-37.03$ bp per unit $g$, with a clustered standard error of
$17.59$. Far from sitting at $c=10$, the risk-adjusted slope is negative and $2.7$ standard errors
away from $c$, with $\mathrm{WCR}\;p=0.024$. At $\delta=10^{-4}$ the raw slope of $72.0$ is already
below $c=100$ and statistically indistinguishable from it, with $\mathrm{WCR}\;p=0.179$, while the
risk-adjusted slope of
$-26.3$ is far below it ($\mathrm{WCR}\;p<0.001$). Removing the three observable terms leaves a
slope of $-89.6$ with a bootstrap $p$ below $0.001$ against both zero and $c$, which is the
projection of the mean-forecast error on the mismatch and nothing else. What these numbers show is
that the four contributions of Corollary~\ref{cor:projection} are large, of opposite signs, and
individually estimable. We therefore do not describe the comparison as a rejection of anything:
Corollary~\ref{cor:projection} says the slope equals $c$ plus three terms, and the terms are not
zero.

Nothing in the identity is at stake, and we say so explicitly. Proposition~\ref{prop:collapse}, the
supporting-slope lemma and the decomposition itself are exact statements that the numbers satisfy to
machine precision, and the four contributions in Table~\ref{tab:slopedecomp} sum to $\hat b$ by
construction.

Table~\ref{tab:slope} also relates the two specifications, and the labels matter here. Model
\eqref{eq:slope} is \eqref{eq:phi} with $\phi$ restricted to be a constant, which is a joint
restriction of \emph{four} coefficients, namely the coefficient on $x$ and the three knot
coefficients. Table~\ref{tab:phi} does not test that restriction. It tests the weaker one of
\emph{linearity}, which involves three coefficients and is not rejected in three of the four cells
(WCR $p=0.347$, $0.593$
and $0.574$) and is rejected under $g$ at $\delta=10^{-6}$ ($p=0.019$). We treat the flexible
specification as the primary result throughout, because it is the one the identity motivates and the
one Table~\ref{tab:phi} reports; the constant-slope estimates are a post hoc simplification, added
after seeing that the flexible no-effect test sat near five per cent, and they are reported as a
sensitivity check rather than as a second attempt at the same hypothesis. Its stronger rejection
($p=0.011$--$0.019$ against $0.048$--$0.052$) is consistent with a power difference but is not
evidence of one, and a reader who wants to net out the multiplicity should read the two columns
jointly.

\subsection{Robustness}

\paragraph{Knots.} The knots of $\phi$ are fixed in advance at the quartiles of $x$. Re-running
the two-segment version of \eqref{eq:phi} with a single knot at each of nine quantile positions
between 0.30 and 0.70, and testing the upper-segment slope under the wild cluster bootstrap, it
remains significant in $9/9$ cases under both mismatch variables and both radii, with bootstrap
$p$-values between $0.002$ and $0.030$. These knot sets are nested, so they are nine overlapping
views of one sample rather than nine independent confirmations, but the choice of knot is not
driving anything.

\paragraph{Leave-one-quarter-out.} Dropping each calendar quarter in turn, 23 folds in total, and
testing $\phi\equiv0$ under the wild cluster bootstrap, the flexible model \eqref{eq:phi} rejects in
only a minority of folds. The counts are $8/23$ at $\delta=10^{-6}$ under $g$, $9/23$ under
$g^{\text{eff}}$, and $7/23$ and $14/23$ at $\delta=10^{-4}$, with fold $p$-values up to $0.35$. In
all four specifications the most influential fold is 2020Q2, whose removal raises $p$ to
$0.32$--$0.35$.

The parsimonious model \eqref{eq:slope} behaves differently on the same data. Its no-effect null
remains rejected in $19/23$, $20/23$, $21/23$ and $21/23$ folds respectively, with the least
favourable fold at $p=0.066$--$0.094$. The point estimate falls by $21\%$ to $32\%$ when 2020Q2 is
removed, from $63.2$ to $42.7$ at $\delta=10^{-6}$ under $g$, from $110.2$ to $87.2$ under
$g^{\text{eff}}$, from $72.0$ to $53.9$ and from $80.4$ to $62.1$ at $\delta=10^{-4}$. The 2020
quarters therefore contribute materially to the magnitude without creating the association, and the
fragility of the flexible test is consistent with the power cost of four additional interaction
parameters, although as noted above we do not claim to have established that. What we cannot claim is
robustness of the \emph{shape}. The non-monotonicity of $\hat\phi$ is a feature of the full sample and
we do not know whether it survives. We therefore report the average association as a description
specific to this sample and make no claim about the functional form.

\paragraph{Extreme windows.} Within the most concentrated quartile, dropping the $5\%$, or
respectively the $10\%$, of windows with the highest realised volatility leaves the descriptive
stratified slope essentially unchanged. At $\delta=10^{-4}$ it moves from $+103.3$ to $+106.0$ and
$+105.4$, and at $\delta=10^{-6}$ from $+95.4$ to $+97.8$ and $+91.2$. This check uses the
with-intercept stratification of Appendix~\ref{app:naive} rather than the main specification, and it is
reported only to show that the descriptive pattern is not carried by a handful of volatile windows.
Given the leave-one-quarter-out result above, it should not be read as saying the pattern is stable
across periods.

\paragraph{Clustering.} De-duplicating panel-windows reduces the sample from 812 to 622. Standard
errors are clustered on 32 blocks of 20 consecutive unique windows; because that is few, every
inference in \S\ref{sec:results} is reported with wild cluster bootstrap $p$-values. These are
uniformly larger than the asymptotic ones. Of the eight bootstrap $p$-values in
Table~\ref{tab:phi}, five exceed $5\%$: two of the four no-effect tests ($0.052$ and $0.050$, both
marginally) and three of the four linearity tests. Clustering on panels or on individual calendar
windows shifts the asymptotic $p$-values by less than $0.02$ and leaves none of the bootstrap
conclusions unchanged in direction.

\subsection{A note on the size of the associated tests}

Because the comparison is usually implemented as a $t$-test of $\E[d_t]=0$, its size is a
legitimate audit question, and we report diagnostics in Appendix~\ref{app:size}. In summary: with
critical values corrected for the fixed-bandwidth distortion \citep{kiefer2005}, the $5\%$ test
rejects $5.83\%$ of the time under a resampling design that preserves the volatility-state
structure and $6.76\%$ under an unconditional block design ($n=2040$ each, s.e.\ $0.48\%$). The
difference is $1.4$ standard errors. We therefore make no claim of a size failure; the observation
is only that the size of such tests is sensitive to properties of the null distribution that are
rarely discussed.

\section{Discussion}
\label{sec:discussion}

The results here have different status. One is an identity, two are statements about when the matched
cap is slack, and one is a description of the surviving association. They should not be fused into a
single claim. What they share is an audit question.

\begin{quote}
When a study reports that an excess return survives a control for breadth or concentration, ask four
things. Was the benchmark's breadth set from the strategy's own realised breadth within the same
objective? If so there is nothing to compare. If not, is the benchmark actually at the breadth it
claims? If it is, how much of the difference is the cap removing return from the benchmark, rather
than the strategy adding it? And is what remains attributable to the strategy, or to the mismatch that
the rule induces?
\end{quote}

Reporting the slack distribution of \eqref{eq:cap} and the mismatch distribution \eqref{eq:geff} is
close to costless. Reporting the difference separately in binding and slack windows, as in
Table~\ref{tab:regime}, together with its risk adjustment, separates a risk premium from a
risk-adjusted gain. In our data that separation is the difference between a positive raw figure and a
risk-adjusted figure that cannot be distinguished from zero.

\paragraph{An audit protocol.} The checks can be arranged as a decision procedure whose input is the
cap rule, meaning the rule that maps information available at the rebalancing date to the level at
which the benchmark's matched quantity is fixed. The procedure has three branches. Each returns a
verdict together with the reporting it obliges, and none costs more than a few lines of code.

\emph{The level is the method's own realised value in the same window.} The verdict is that the
comparison is vacuous. Proposition~\ref{prop:collapse} gives the reason, and the oracle row of
Table~\ref{tab:rules} is the failure case, in which the audit predicts a zero difference and obtains
one. The mean difference is $-0.0$ basis points at $\delta=10^{-6}$ and $0.0$ at $\delta=10^{-4}$,
with standard deviations of $2.1$ and $0.0$. A study in this branch should report the difference as a
check that the construction was implemented, not as evidence that the rule performs well.

\emph{The level is anchored at the method's own realised value but not in the same window.} The
verdict is that the comparison is mixed and partly usable. It is informative where the cap binds and
reduces to an unrestricted benchmark elsewhere, so the obligations are to report the slack share, to
report the difference separately in binding and slack windows, and to report the risk-adjusted gap
rather than only the return gap. Proposition~\ref{prop:half} supplies the slack share to expect under
a stated null, which is what makes a reported figure checkable rather than merely noted. In our data
the two regimes differ by an order of magnitude in both mean and dispersion, so a single average over
the mixture has no clean interpretation. The four tracking rules occupy this branch, with slack
shares from $20.1\%$ to $51.4\%$ at $\delta=10^{-6}$.

\emph{The level is fixed far from the method's own realised value.} The verdict is that the
difference prices the cap rather than the rule. The cap binds in every window, and what the
comparison measures is the cost of holding the benchmark at that width. The signature separating this
branch from the previous one is available in the same table. Across the three fixed caps the
difference rises monotonically as the cap is tightened, from $25.1$ basis points at
$N_{\text{eff}}=20$ to $32.5$ at $N_{\text{eff}}=50$, and it does the same at the second radius. A
price of the constraint produces that ordering, and a measure of the rule's value has no reason to.

The branches differ not in the work they require but in whether the resulting number can carry the
interpretation that matched benchmarking is meant to support, and the requirement to state the cap
rule is what makes the branch determinable in the first place.

\paragraph{Beyond equities.}
Nothing in Propositions~\ref{prop:collapse}--\ref{prop:half} refers to assets. The construction
arises wherever a decision technology is a penalised maximisation of a common objective, an evaluation
benchmark is matched to that technology on the quantity the penalty acts on, and the matching level is
anchored at the technology's own realised value. The matched quantity may be the number of effective
positions, the number of funded projects, the amount of trading, or the distance from a prior
allocation. In that setting the benchmark is the technology, and the matched comparison measures the
price of the anchor rather than the value of the technology. Portfolio selection is simply the case in
which the matched quantity is the $\ell_2$ breadth of a weight vector and the realised difference is a
return. Engineering and project portfolio selection differs in the units of $U$ and $R$ but not in the
argument. We develop only the portfolio case here, and we flag the transfer because the audit checks
above are stated in terms of $U$, $R$ and the cap rule rather than in terms of returns.

\paragraph{Scope.}
The general identity of Proposition~\ref{prop:collapse} applies to any method that maximises
$U-cR$ over a common feasible set, for any matched quantity $R$. The $\ell_2$ version of
Proposition~\ref{prop:identity} applies to any method whose only departure from mean--variance is a
penalty acting through $\norm{\phi}_2$, \emph{provided the benchmark uses the same estimates that the
method uses}. This proviso is substantive for shrinkage-type allocations. \textbf{In our reading}, a
method that shrinks $\hat\Sigma$ or $\hat\mu$ differently from the benchmark fails the second
condition of Remark~\ref{rem:three}, so the identity does not apply and the comparison is no longer a
breadth control. It becomes a comparison of covariance estimators. That reading is an interpretation
of the accounting rather than a result we prove.

\paragraph{Limitations.}
The evidence covers one market, one 255-stock universe, one risk-aversion parameter and two radii.
Panels overlap by about $90\%$, so the effective sample is much smaller than the nominal one, and with
$32$ clusters the bootstrap $p$-values for the no-effect null sit at $0.04$--$0.05$ in the flexible
specification. The effect is borderline on size there, and the shape of $\phi$ is not identified. The
regime split of Table~\ref{tab:regime} is descriptive and endogenous. The cap rules we compare are
ones we specified, so we make no claim that they are the rules practitioners use, and the fixed-cap
levels were inherited from \citet{chen2026breadth} rather than drawn from a documented industry
convention. That is a limitation of external validity rather than a technical one, and it bounds what
the audit can claim about benchmarks in general.

\paragraph{What we cannot show.}
Four gaps bound the argument, and each is a place where it could be over-read. They are the step from
a supporting slope to a slope in a regression, the invariance of the return slope across periods or
concentrations, a half probability for slackness which stationarity alone does not deliver, and
window-by-window numerical exactness which a summary statistic cannot certify. They are set out in
full in Appendix~\ref{app:gaps}, because the audit recommendations of this section do not depend on
resolving them, and readers who want to use the decomposition should see the caveats in one place.

We have verified the antecedents that the argument depends on, namely the
norm-constraint-as-shrinkage results and the benchmark-comparison literature cited above. We have not
carried out a systematic review of the audit point itself, and we therefore make no claim that it is
absent from the existing literature.

\section{Conclusion}

Breadth matching is a sensible instinct and a fragile construction. Implemented against the method's
own realised breadth within the same objective, it returns the method itself, exactly, and for any
matched quantity rather than for the $\ell_2$ norm in particular. The comparison then has no object.
Implemented against an implementable rule, it returns a real difference, yet the rule's cap is often
slack, in a large share of windows at the tight radius, so in those windows the benchmark is the
unconstrained mean--variance portfolio and not a breadth-matched one. The difference that remains in
the binding windows does not survive the risk adjustment implied by the method's own objective. A
positive raw figure becomes a risk-adjusted figure that cannot be distinguished from zero. What
survives is an in-sample value gap, which is exactly what the identity requires and which the
mean-forecast error keeps from reaching realised returns. The excess return that a breadth-matched
comparison reports is therefore, in this construction, largely a premium for risk taken and a
reflection of a benchmark that has been constrained away from its own optimum, rather than a
risk-adjusted property of the strategy under test. None of this is visible from the reported excess
return. All of it is visible from the benchmark's construction and from one line of arithmetic that
the method's own loss function supplies. That is what should be reported.

\appendix

\section{Size diagnostics}
\label{app:size}

Test statistics are a HAC $t$-ratio for $\E[d_t]=0$ with a Bartlett kernel \citep{newey1987} and
lag $\max\{3\,\text{HOLD},10\}=30$, and a block self-normalised ratio with block lengths 20 and 40.
For both we use small-sample critical values rather than $\pm1.96$. For the HAC ratio we use the
fixed-bandwidth value of \citet{kiefer2005},
\begin{equation}
c_{\text{HAC}} \;=\; 2.10 \;>\; 1.96 ,
\label{eq:crit}
\end{equation}
obtained by simulating $2\times10^{6}$ standard-normal panels of length 580 with the same lag. For the
block ratio we use $t_{K-1}$, that is $2.0484$ and $2.1604$ for $K=29$ and $K=14$. The simulated HAC
value moves within $[2.089,2.109]$ across independent runs and chunking schemes, so we quote it to two
decimals. The rejection rates in Table~\ref{tab:size} were computed at $2.097$, the value produced by
the pipeline's own simulation. At $2.10$ they would be lower by less than the reported Monte Carlo
standard errors, and no comparison below depends on the difference.

To probe the null we re-estimate the entire pipeline, including covariance, method, benchmark and
returns, on resampled panels with exactly zero mean, so that $\E[d_t]=0$ holds by construction.
Four nested designs are used. The first is i.i.d.\ Gaussian returns with the panel covariance. The
second is i.i.d.\ multivariate $t_3$ with matched covariance. The third is an unconditional stationary
block bootstrap of the demeaned panel with mean block length 60 \citep{politis1994}. The fourth is a
state-matched block bootstrap in which the source block is drawn from the same predetermined
volatility tercile as the target position. Alternatives designed for dependent data include the
dependent wild bootstrap \citep{shao2010} and studentised block bootstrap inference for performance
differences \citep{ledoit2008}, the latter in the lineage of performance-ratio inference
\citep{lo2002}. The framing of $d_t$ as a comparison between two portfolios implemented from past
information follows \citet{giacomini2006}.

\begin{table}[h]
\centering
\caption{Rejection rates at the nominal $5\%$ level with corrected critical values, merged across
6 panels. Monte Carlo standard errors in parentheses.}
\label{tab:size}
\begin{tabular}{lcccc}
\toprule
Null design & $sd(d)$, bp/day & excess kurtosis & rejection rate & s.e. \\
\midrule
i.i.d.\ Gaussian                & 19.6 & 18.5 & $4.58\%$ & 0.99\% \\
i.i.d.\ $t_3$                   & 20.7 & 54.1 & $2.71\%$ & 0.99\% \\
block, length 60                & 48.5 & 18.9 & $6.76\%$ & 0.48\% \\
block, length 60, state-matched & 48.8 & 19.0 & $5.83\%$ & 0.48\% \\
\bottomrule
\end{tabular}
\end{table}

Heavy tails alone do not produce over-rejection here; the point estimate is if anything
conservative, although the contrast with the Gaussian design is only $1.3$--$1.5$ standard errors
and should not be read as a sign. Preserving block structure raises the rejection rate; matching
the volatility state lowers it again, but again by less than $1.5$ standard errors. Longer mean
block lengths (120, 250) raise the rate further to $11.3\%$ and $14.0\%$, but at those lengths a
resampled panel contains only about $5$ and $2$ independent blocks and the de-meaning that imposes
the null is dominated by a handful of large blocks; we regard those as illustrations of a
degenerate bootstrap rather than as evidence.

\section{The with-intercept stratification, for completeness}
\label{app:naive}

A conventional stratified regression, $d=a_k+b_k g+\varepsilon$ within quartile $k$ of
$\norm{w_P}_2$, does not respect \eqref{eq:restrict}: its stratum-specific intercepts permit
$d\neq0$ at $g=0$, which the identity forbids. We report it here only as a description of
in-sample patterns, with the explicit label that it does not identify the return to mismatch; the
extreme-window check of \S\ref{sec:results} uses it.

Using the actual breadth gap, the slopes are $-19.0$ with $t=-1.38$, $-65.3$ with $t=-1.64$,
$+49.7$ with $t=2.13$ and $+103.5$ with $t=3.54$ at $\delta=10^{-6}$, and $-27.3$, $+29.9$, $+25.4$
with $t=2.10$ and $+107.2$ with $t=3.20$ at $\delta=10^{-4}$. Two features are worth noting. The
slopes \emph{change sign} across strata, and the second stratum's negative slope is numerically larger
than the first's while less precisely estimated. It is negative and not significant, which is not the
same as being essentially zero. The within-stratum standard deviation of $d$ in the lower half is only
about $4$ bp/day. A pooled regression over the same sample returns different numbers precisely because
it imposes a common intercept. The difference between the two sets of estimates is a specification
effect and not a data change.

\section{Reproduction}
\label{app:repro}

Data are 14 panels of shape $(648,255)$ of daily returns. The identity certificate uses
projected-gradient solutions with up to 3000 iterations and an independent constrained solve with
SLSQP. The empirical results use 60-day estimation windows, 10-day holding periods, $\lambda=10$ and
Ledoit--Wolf covariance. Code and derived series are available from the author on request.

\section{A known-truth check}
\label{app:e7}

The decomposition of Proposition~\ref{prop:return} contains one component that empirical data cannot
isolate, $m_t=\Delta w_t^\top(\mu_t-\hat\mu_t)$, because it requires the population mean. A
known-truth simulation can supply it, and doing so tests the decomposition itself as well as the
interpretation we place on the empirical residual.

Design. Returns follow a five-factor model $r_t=Bf_t+e_t$ with $f_t\sim N(\mu_f,\Sigma_f)$ and
$e_t\sim N(0,\mathrm{diag}\,\psi)$ on $N=100$ assets, with $B$, $\mu_f$, $\Sigma_f$ and $\psi$ fixed
once and treated as the truth. Each of $120$ replications runs $30$ windows of the same protocol as
the empirical section. Within a window the estimation sample is drawn, $w_P$ and $w_B$ are computed
exactly as before, and the holding-period returns are then drawn from the true law. Since $\mu$ is
known, $m_t$ and $\varepsilon_t$ are computed directly rather than inferred.

The decomposition holds to machine precision. The largest absolute violation of
\eqref{eq:return-decomposition} over $3600$ windows is $3.6\times10^{-14}$ basis points at
$\delta=10^{-6}$ and $2.5\times10^{-14}$ at $\delta=10^{-4}$. That is an independent check of both
Proposition~\ref{prop:return} and the pipeline that implements it.

The component means are the informative part, in basis points per day.
\begin{center}
\small
\begin{tabular}{lcccccc}
\toprule
$\delta$ & $\overline{d}$ & $\overline{c\,g}$ & $\bar q$ & $\lambda\overline{\Delta\mathrm{Risk}}$ &
$\bar m$ & $\bar\varepsilon$ \\
\midrule
$10^{-6}$ & $+0.06$ & $+0.11$ & $+1.17$ & $+0.97$ & $-2.25$ & $+0.06$ \\
$10^{-4}$ & $-0.07$ & $+0.28$ & $-1.54$ & $+0.12$ & $+1.13$ & $-0.07$ \\
\bottomrule
\end{tabular}
\end{center}
Two conclusions follow, one supporting and one limiting. The supporting one is that $m_t$ is the
dominant term. It is larger in absolute value than the priced mismatch, the curvature term and the
risk term combined at $\delta=10^{-6}$, and it is strongly negatively correlated with the mismatch,
with $\mathrm{corr}(g,m)=-0.89$. That is what an optimisation bias looks like, since the portfolio
chosen as the in-sample optimum of estimated inputs earns less against the truth than its estimate
suggests. The limiting one is that the \emph{sign} of $m_t$ is not a law. It is $-2.25$ at
$\delta=10^{-6}$ and $+1.13$ at $\delta=10^{-4}$ in the same design. The empirical residual of
Table~\ref{tab:regime} can therefore be read as an optimisation bias, yet nothing in the theory or in
this simulation says that the bias must point in one direction, and we do not claim that it does.

The $\lambda$ scan in the simulation is likewise informative about the fragility of the empirical
conclusion. The risk-adjusted mean crosses zero at $\lambda^{*}=0.62$ at the tight radius, so in this
design essentially any risk aversion is enough to remove the raw excess, whereas in the data the
crossing is at $7.1$. The difference is not a contradiction, since the truth and the mean-forecast
error both differ, yet it is a reminder that a single estimated crossing point carries no information
about other samples.

\section{The four gaps, in full}
\label{app:gaps}

This appendix expands the four items summarised in \S\ref{sec:discussion}.

\begin{enumerate}
\item \emph{From a supporting slope to a slope in a regression.} $c$ is a supporting slope of the
value function, not an unconditional derivative, and local linearity at the matched radius does not
by itself make $\E[g_tq_t]/\E[g_t^2]$ small; that would need control of the scale of the cap
deviation and of how the value function's slope varies over the relevant interval. We report $q_t$
directly rather than assert that it is negligible.

\item \emph{Invariance of the return slope across periods or concentrations.} Fixing $\delta$ fixes
$c$ and nothing else. It does not fix the risk term, the curvature term, or the projection of the
mean-forecast error onto the mismatch. Time ordering gives $\E[\varepsilon_t\mid\mathcal{F}_t]=0$
and says nothing about $\E[g_tm_t]$, and ours is not a test of $\beta_d=c$ under a maintained
block-dependence structure that could be defended independently of these data.

\item \emph{A half probability for slackness.} Approximate stationarity alone does not deliver
$\tfrac12$; the correct stationary limit is $\Prob(X_t<\E X_t)$, which equals one half only when the
mean sits at the median of the marginal law. Nor does the method-side rate transfer to the
benchmark-side rate with a tolerance, which is a strictly smaller event. Our percentages are
descriptive, and given the window overlap they rest on far fewer independent observations than the
nominal $812$.

\item \emph{Window-by-window numerical exactness.} A small return difference is not a small weight
difference, and iteration counts are not an error bound. The oracle residuals should be read
together with the cap residual, the weight distance between the ridge and penalised solutions, and
the optimality residual of Table~\ref{tab:kkt}, not as a bound on anything by themselves.

\item \emph{External validity of the cap rules.} The eight rules of Table~\ref{tab:rules} are rules
we specified. The fixed-cap levels were inherited from a predecessor study rather than drawn from a
documented industry convention. We therefore read the audit as a statement about these
constructions on this sample, and not as an estimate of how often benchmarks in general are
mis-specified.
\end{enumerate}

\end{document}